\documentclass[fleqn,usenatbib]{mnras}

\usepackage{newtxtext,newtxmath}

\usepackage[T1]{fontenc}
\usepackage{ae,aecompl}

\usepackage{graphicx}	
\usepackage{amsmath}	
\usepackage{siunitx}
\usepackage{booktabs}
\usepackage{multirow}
\usepackage{color, colortbl}
\usepackage{mathrsfs}
\usepackage[normalem]{ulem}
\usepackage{lineno}
\usepackage{xcolor}
\usepackage{xspace}
\usepackage{graphicx}

\let\vec\mathbf

\newcommand{\stkout}[1]{\ifmmode\text{\sout{\ensuremath{#1}}}\else\sout{#1}\fi}

\usepackage{ulem}
\newcommand{\redmapper}{redMaPPer\xspace}
\DeclareMathOperator*{\zbcg}{\textit{z}_{\mathrm{BCG}}\xspace}

\DeclareMathOperator*{\zmem}{\textit{z}_{\mathrm{mem}}\xspace}
\DeclareMathOperator*{\lstar}{\textit{L}_\star\xspace}
\DeclareMathOperator*{\fcl}{\textit{f}_{\text{cl}}\xspace}
\DeclareMathOperator*{\fproj}{\textit{f}_{\text{proj}}}
\DeclareMathOperator{\lambdaobs}{\lambda\xspace}
\DeclareMathOperator{\lambdatrue}{\lambda_{\text{spec}}\xspace}

\DeclareMathOperator{\deltaznormed}{\Delta z / (1+z)\xspace}

\DeclareMathOperator*{\pmem}{\textit{p}_{\mathrm{phot}}\xspace}

\DeclareMathOperator*{\pspec}{\textit{p}_{\mathrm{spec}}}

\DeclareMathOperator*{\pred}{\textit{p}_{\mathrm{red}}}

\title[Spectroscopic Quantification of Projection Effects]{Spectroscopic Quantification of Projection Effects\\ in the SDSS \redmapper\ Galaxy Cluster Catalogue}

\author[Myles et al.]{
\parbox{\textwidth}{
\Large
J.~Myles$^{1,2,3}$,\thanks{E-mail: jmyles@stanford.edu}
D.~Gruen$^{1,2,3}$,
A.~B.~Mantz$^{1,3}$,
S.~W.~Allen$^{1,2,3}$,
R.~G.~Morris$^{1,2,3}$,
E.~Rykoff$^{2,3}$,
M.~Costanzi$^{4,5,6}$,
C.~To$^{1,2,3}$,
J.~DeRose$^{7,11}$,
R.~H.~Wechsler,$^{1,2,3}$,
E.~Rozo$^{12}$,
T.~Jeltema$^{10,11}$,
E.~R. Carrasco$^{13}$,
A.~Kremin$^{8}$,
R.~Kron$^{9}$
}
\vspace{0.4cm}
\\
\parbox{\textwidth}{
(Affiliations are listed at the end of the paper)
\vspace{0.8cm}
}
}

\date{Accepted 2021 April 26. Received 2021 April 26; in original form 2020 November 18}

\pubyear{2020}
\begin{document}
\label{firstpage}
\pagerange{\pageref{firstpage}--\pageref{lastpage}}
\maketitle

\begin{abstract}
Projection effects, whereby galaxies along the line-of-sight to a galaxy cluster are mistakenly associated with the cluster halo, present a significant challenge for optical cluster cosmology. We use statistically representative spectral coverage of luminous galaxies to investigate how projection effects impact the low-redshift limit of the Sloan Digital Sky Survey (SDSS) \redmapper galaxy cluster catalogue. Spectroscopic redshifts enable us to differentiate true cluster members from false positives and determine the fraction of candidate cluster members viewed in projection. Our main results can be summarized as follows: first, we show that a simple double-Gaussian model can be used to describe the distribution of line-of-sight velocities in the \redmapper sample; second, the incidence of projection effects is substantial, accounting for $\sim 16$ per cent of the weighted richness for the lowest richness objects; third, projection effects are a strong function of richness, with the contribution in the highest richness bin being several times smaller than for low-richness objects; fourth, our measurement has a similar amplitude to state-of-the-art models, but finds a steeper dependence of projection effects on richness than these models; and fifth, the slope of the observed velocity dispersion--richness relation, corrected for projection effects, implies an approximately linear relationship between the true, three-dimensional halo mass and three-dimensional richness. Our results provide a robust, empirical description of the impact of projection effects on the SDSS \redmapper cluster sample and exemplify the synergies between optical imaging and spectroscopic data for studies of galaxy cluster astrophysics and cosmology.
\end{abstract}

\begin{keywords}
galaxies: clusters: general -- galaxies: groups: general
\end{keywords}



\section{Introduction} 
\label{sec:intro}
The predicted number density of galaxy clusters as a function of mass and redshift, known as the cluster mass function, is sensitive to the parameters describing the underlying cosmological model. As a result, measurements of cluster number counts have been used to place competitive constraints on the mean matter density of the Universe, $\Omega_m$, the dark energy density, $\Omega_{\rm DE}$, and equation-of-state parameter, $w$, as well as to provide complementary information on modified gravity and inflation \citep[e.g.][and references therein]{Allen1103.4829, Mantz2015, Planck2018a, Bocquet2019, Costanzi2019}. Galaxy clusters stand out among cosmological probes due to their ability to be readily detected and studied in detail across a broad range of wavelengths, with each wavelength offering certain complementary advantages. 

Optical galaxy surveys identify clusters by finding overdensities of galaxies on the sky. Relative to other wavelengths, optical surveys are especially sensitive to lower mass clusters, for which X-ray photons from the virialized gas in the intracluster medium can be scarce and the Sunyaev-Zel'dovich effect signal is small. Another advantage of optical measurements is that they can be extracted from the same imaging surveys being used for a broad range of other cosmological experiments, facilitating self-consistency tests and constraint combinations. Critically, optical imaging surveys in well-selected filters also provide robust photometric redshifts for the clusters found, as well as weak gravitational lensing measurements that can be used to calibrate absolute cluster masses \citep[e.g.][]{WTGIII, WTGI, Hoekstra2015, Simet2017, Schrabback2018, Dietrich2019, McClintock2019}. Because optical cluster-finding straightforwardly extends down to lower mass clusters, it can be especially valuable for extracting cosmological information from the clustering of galaxy clusters \citep{mana2013, to2020a, to2020}, as well as certain tests of modified gravity \citep{Cataneo2018b, Cataneo2018a}.

The primary challenge in using galaxy clusters to test cosmological models is to determine the relatively complex relationship between the observable properties and mass, recalling that it is the mass function, rather than the number of clusters as a function of observables, that is directly predicted by theory. For optical clusters, one such observed property is the cluster \textit{richness}, which commonly refers to the number of galaxies associated with a given cluster. For optically selected clusters, this amounts to determining the relationship between richness and mass. The \textbf{red} sequence \textbf{Ma}tched filter \textbf{P}robabilistic \textbf{Per}colation (\redmapper) algorithm --- a leading method for detecting and measuring galaxy clusters with optical imaging survey data --- addresses this challenge by leveraging the red sequence of galaxies to infer photometric redshifts, and thus determine a probabilistic assignment of member galaxies to clusters. By defining \redmapper richness, $\lambda$, to be the probability-weighted count of \textit{bright red} cluster members, rather than all detected and selected member galaxies, the algorithm is able to reduce the otherwise prohibitive scatter in the mass--richness relation. 

Even the \redmapper cluster richness, however, is subject to \textit{projection effects}, whereby galaxies along the line-of-sight to a cluster are incorrectly classified as cluster members \citep{redmapper4, Costanzi1807.07072, Sohn1712.00872, Sunayama2020}. This systematic error is inherent to finding clusters with imaging data due to the limited precision of standard photometric observables.\footnote{The photometric redshift error of $\sigma_z \approx 0.006$ at $z \approx 0.1$ cited in \cite{redmapper1} corresponds to a line-of-sight co-moving distance of $\approx 24.5$ Mpc, which in turn corresponds to a physical distance of $d = \frac{\chi}{1+z} \approx 22$ Mpc.}. Ideally, projection effects would be calibrated precisely with hydrodynamical simulations that describe the formation and evolution of galaxies within their host dark matter haloes. However, this goal is infeasible at present due to the complex challenges involved in simulating cluster galaxy evolution.

In the absence of robust, quantitative predictions for the impact of projection effects, we must turn to empirical methods. Observationally, there are two main routes to quantify projection effects: utilizing X-ray observations of the intracluster medium and optical spectroscopy of cluster member galaxies. X-ray observations of the virialized hot gas within clusters are essentially unaffected by projection effects, due to the nearly unique association of extended X-ray emission with virialized cluster gas, and the density-squared dependence of this emission. Quantities such as the X-ray emitting gas mass, gas temperature, and their product, $Y_x$, are tightly correlated with the three-dimensional halo mass, enabling precise determinations of the shape and intrinsic scatter of mass--observable scaling relations \citep{reiprich2002, Mantz2010a, Mantz2010b, Mantz2010c, Mantz2015, schellenberger2017, erosita1}. Complementing such measurements, optical spectroscopy of candidate cluster members identified by optical imaging surveys offers a way to determine which galaxies lie within the virialized three-dimensional halo structure \citep{redmapper4, Sohn1712.00872}. Here, the challenge is to obtain sufficient, representative coverage of the typically tens of bright galaxies determined by \redmapper to belong with some probability to each of a large sample of clusters. 

In this paper, we report on our use of optical spectroscopy to characterize projection effects in an optically selected cluster sample, using the \redmapper cluster catalogue constructed from the Sloan Digital Sky Survey (SDSS) DR8 imaging data \citep{sdss_dr8}. By making use of the extensive optical spectroscopy of candidate cluster members also available from SDSS, we are able to differentiate true member galaxies from objects in projection, and derive a robust, quantitative determination of the impact of projection effects on observed richness as a function of richness. We compare our empirical results to state-of-the-art models of projection effects and quantify the differences. We comment on the potential significance of our findings for recent cluster cosmology results from the Dark Energy Survey, which reported a $5.6 \sigma$ tension in the $\sigma_8-\Omega_m$ parameter plane with respect to the \textit{Planck} primary CMB analysis \citep{desy1}.

Our paper is organized as follows. In Section \ref{sec:data}, we describe the data used to constrain the impact of projection effects in \redmapper galaxy clusters. In Section \ref{sec:formalism}, we describe our formalism for modeling projection effects. We present our empirical measurement of projection effects in Section \ref{sec:modeling}, and compare our measurement to analogous predictions from both simulations and analytic modeling in Section \ref{sec:sims}. The implications of our results are discussed in Section \ref{sec:discussion}. We conclude and provide suggestions for future work in Section \ref{sec:conclusion}. A flat $\Lambda$CDM cosmology with $H_0= 70 $ km s$^{-1}$ Mpc$^{-1}$ and $\Omega_m = 0.3$ is assumed throughout.

\section{Data}
\label{sec:data}

The Sloan Digital Sky Survey (SDSS) provides a three-dimensional map of the sky produced from a combination of imaging and spectroscopic data collected with a dedicated 2.5-m telescope at Apache Point Observatory. Here we use the SDSS Data Release 8 (DR8) galaxy catalogue \citep*{sdss_dr8}, which consists of 14,000 deg$^2$ of drift-scan imaging in the northern and southern Galactic caps. After quality cuts have been applied to the data, 10,500 deg$^2$ of imaging remains. The corresponding spectroscopic catalogue contains over 1.3 million spectroscopic redshifts \citep*{sdss_spec}.

For our galaxy cluster sample, we use the SDSS DR8 \redmapper cluster catalogue, which identifies cluster candidates as overdensities of red sequence galaxies on the sky \citep*{redmapper1}. The \redmapper catalogue takes as input a catalogue of galaxy fluxes and positions, and generates as output a catalogue of galaxy clusters. \redmapper computes the probability (labelled $\pmem$ in this work) of each input galaxy to be a member of each cluster and the probability of each input galaxy to be the \redmapper Bright Central Galaxy (BCG) of its associated cluster. The \redmapper algorithm is currently a leading tool for identifying galaxy clusters from optical imaging data, having demonstrated superior performance in delivering a relatively low-scatter richness measure and precise photometric redshifts, as verified by \citet{redmapper1, redmapper2, redmapper3, redmapper4}. 

The SDSS DR8 \redmapper cluster catalogue contains $\sim 25,000$ clusters from $0.08 \leq z \leq 0.55$. Spectroscopic redshift measurements exist for a subset of this catalogue from the SDSS Legacy and Special programs and the SEGUE-1 and SEGUE-2 surveys \citep{segue1, sdss_dr8, sdss3}.
Statistically representative spectroscopic coverage is available for galaxies brighter than $m_r = 17.77$. For this study, we use SDSS DR8 \redmapper clusters with photometric redshifts $0.08 \leq z \leq 0.12$, which results in a limiting galaxy luminosity for spectroscopic follow-up of $L \geq 0.55 \lstar$ in the SDSS \textit{i}-band. Past analyses suggest a spectroscopic redshift failure rate of $~0.9$ per cent for this sample. \citep{redmapper4}. We make our measurements in six richness bins, restricting the analysis to clusters with $\lambda \geq 5$. While past cosmological analyses have used $\lambda \geq 20$ as a cut, we choose $\lambda \geq 5$ as a lower threshold to most effectively test for any richness dependence in the data. We measure galaxy velocities relative to the \redmapper BCG, and thus restrict our sample to clusters for which the BCG identified by \redmapper has a measured spectroscopic redshift. The numbers of clusters and candidate member galaxies with spectroscopic measurements are shown in Table~\ref{tab:sdss_data}. We illustrate the completeness of the spectroscopic coverage of our cluster member sample in Appendix \ref{app:completeness}.

\begin{table*}
\begin{center}
\begin{tabular}[h]{ c c c c c c c c } 

& Richness & 5 -- 20 & 20 -- 27.9 &  27.9 -- 37.6 & 37.6 -- 50.3 & 50.3 -- 69.3 & 69.3 -- 140\\ 

  \hline
  \hline
\multirow{2}{*}{$L \geq 0.55 \lstar$} & Clusters   & 2569 & 149 & 69 & 38 & 17 & 7\\ 
& Members & 11547 & 1575 & 1032 & 712 & 441 & 249\\ 

\rule{0pt}{4ex} 

\multirow{2}{*}{$0.55 \leq \lstar < 0.9$} & Clusters   & 2256 & 145 & 68 & 38 & 17 & 6\\
& Members & 5845 & 802 & 508 & 346 & 227 & 121\\

\rule{0pt}{4ex} 

\multirow{2}{*}{$L > 0.9 \lstar$} & Clusters   & 2209 & 148 & 69 & 38 & 17 & 7\\
& Members & 5702 & 773 & 524 & 366 & 214 & 128\\

\end{tabular}\caption{Number of clusters and non-BCG members with spectroscopic redshifts in each richness bin of each sub-sample analysed. For this analysis, we assume the most likely BCG is the true BCG.}\label{tab:sdss_data}
\end{center}
\end{table*}

\section{Model Formalism}
\label{sec:formalism}

The complex astrophysics of galaxy evolution makes direct predictions of the mass--richness relation for optically selected galaxy clusters very challenging. The problem can become more tractable, however, by forming an intermediate observable or observables that can be expected to correlate more tightly with mass than the observed, photometric richness, $\lambdaobs$. Here, we use optical spectroscopy of candidate \redmapper cluster members, meaning galaxies that \redmapper has identified as associated with a cluster, to identify galaxies that are physically associated with the virialized cluster haloes. We form the intermediate mass proxy, $\lambdatrue$, a modified richness estimate informed by both photometric and spectroscopic data, that can be expected to exhibit reduced systematic scatter at fixed mass. This approach is analogous to the use of low scatter X-ray mass proxies to supplement survey observables within cluster counts analyses (e.g.\ \citealt{Vikhlinin_2009, Mantz2010a, Benson_2013}).

\subsection{Modeling galaxies in projection}
\label{sec:measuref}

\begin{figure}
    \begin{minipage}{0.45\textwidth}
        \includegraphics[width=0.9\textwidth,scale=0.5]{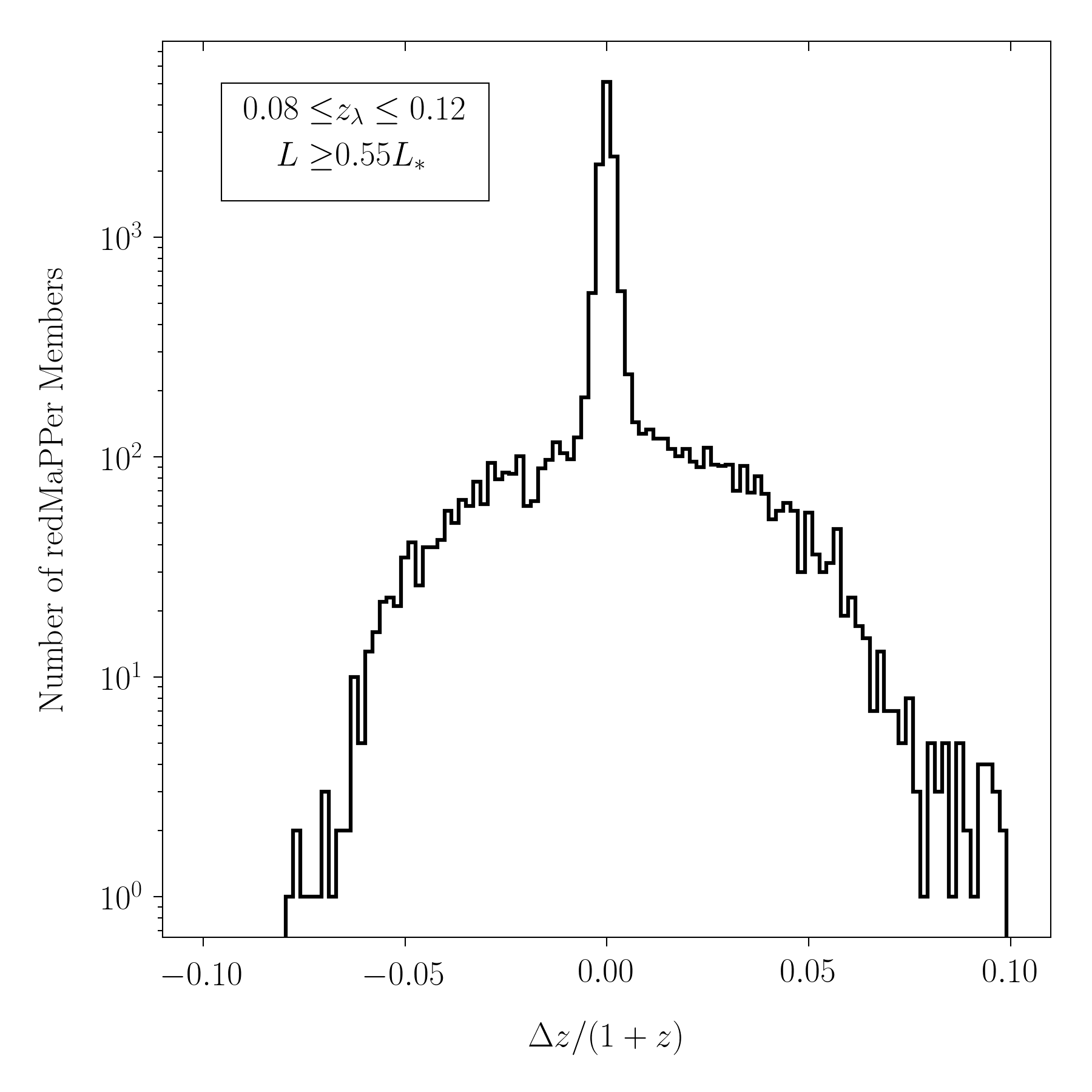}
        \caption{The line-of-sight velocity distribution of candidate redMaPPer members. Two clearly discernible components are observed, corresponding to true cluster halo members and galaxies in projection, respectively. Fifteen (out of 15571) outlying galaxies with $\left|\frac{\Delta z}{1+z}\right| > 0.1$ were identified and removed from all subsequent analysis.}\label{fig:sdss_data}
    \end{minipage}
\end{figure}

We wish to measure the fraction ($\fcl$) of candidate members of a \redmapper cluster of galaxies that are within the virialized halo of the cluster. Equivalently, we measure the fraction, $\fproj \equiv 1 - \fcl$, of candidate cluster members that are viewed in projection along the observer's line-of-sight. 

We begin by proposing a functional form for the distribution of galaxy velocities with respect to the cluster BCG:

\begin{equation}
\deltaznormed = \frac{\zmem - \zbcg}{1 + \zbcg}.
\end{equation}

Note that the BCG spectroscopic redshift, rather than the cluster photometric redshift, must be used here, because the photometric redshift error of $\sigma_z\approx 0.006$ corresponds to a velocity width of $\sim 1800 \text{ km s}^{-1}$, which exceeds the velocity dispersions of low richness clusters. 
We model this distribution as a mixture of two Gaussian distributions, representing galaxies that are true cluster members and those that appear in projection:

\begin{eqnarray} \label{eqn:likelihood}
    p(\deltaznormed) &=& f_{\mathrm{cl}} \ \mathcal{N}(\deltaznormed| \mu_{\mathrm{cl}}, \sigma_\mathrm{cl}) \\
    && + \, f_{\mathrm{proj}} \ \mathcal{N}(\deltaznormed| \mu_{\mathrm{proj}}, \sigma_{\mathrm{proj}}), \nonumber
\end{eqnarray}
where $\mathcal{N}(x|\mu|\sigma)$ represents the density of a normal distribution with mean $\mu$ and standard deviation $\sigma$ evaluated at $x$.
The virial theorem motivates describing the cluster one-dimensional velocity distribution as Gaussian, while the applicability of the overall model is addressed empirically in Section \ref{sec:modeling}.

The $\fproj$ parameter in this model can be straightforwardly interpreted as the fraction of candidate cluster members that are not, in fact, associated with the cluster. \footnote{As discussed by \citet{Farahi2016}, galaxies with line-of-sight velocities consistent with cluster membership cannot be assumed to belong to the \textit{primary} dark matter halo given some prescribed definition for halo membership.} Given the paucity of optical spectroscopic measurements, however, we must constrain this parameter in bins of cluster richness rather than on a cluster-by-cluster basis. We choose the same richness bins as \cite{Costanzi2019}, which were constructed to be populous enough for cosmological cluster counts analysis.

Additionally, the spectroscopic redshift measurements must be unbiased with respect to cluster membership, given the initial selection of possible members.
This is ensured by restricting our study to galaxy luminosities, $L \geq 0.55 \lstar$ in the SDSS \textit{i}-band, above which SDSS spectroscopic coverage is statistically representative over the $0.08<z<0.12$ redshift range, regardless of color (see Section \ref{sec:data}). 

In Appendix \ref{app:validation}, we show that the cluster component has a mean $\deltaznormed$ value consistent with zero for all richness bins, and that the parameters of the projected distribution are consistent with being constant across richness bins.
Our baseline model therefore includes common values of $\mu_\mathrm{cl}$ (fixed to zero), $\mu_\mathrm{proj}$ and $\sigma_\mathrm{proj}$, while the remaining parameters are allowed to be independent in different richness bins ($f_{\mathrm{cl},j}$, $f_{\mathrm{proj},j}$ and $\sigma_{\mathrm{cl},j}$ for the $j$th bin). 

To account for sample variance, we bootstrap over clusters in each richness bin, finding the maximum likelihood parameter values for each bootstrap data set; these parameter distributions are then used to estimate best fitting values and uncertainties. The final best fitting parameter values are shown in Appendix \ref{app:validation}.
For a given set of parameter values, the probability that a candidate member galaxy with a line-of-sight velocity $\deltaznormed$ in richness bin $j$ belongs to the cluster component is
\begin{equation}
\label{eqn:halo_model}
    \pspec \equiv \frac{f_{\mathrm{cl}, j}  \ \mathcal{N}(\deltaznormed| 0, \sigma_{\mathrm{cl},j})}{p(\deltaznormed)},
\end{equation}
where the denominator is given by Equation~\ref{eqn:likelihood}.

For each cluster, we assume the most likely BCG is the true BCG, and we define $\pspec$ as unity for this galaxy.

\subsection{Modeling the richness bias due to projection effects}
\label{sec:lambdabias}

\redmapper, as a red sequence based cluster finder, is designed to identify overdensities of \textit{red} galaxies from photometric data. In constructing an improved richness estimate for \redmapper clusters with spectroscopic follow-up, $\lambdatrue$, we incorporate both photometric and spectroscopic information. Ideally, our new richness definition should be defined to achieve two goals: first, $\lambdatrue$ should be similar enough to $\lambdaobs$ that a comparison of the two serves as a measurement of the extent to which \redmapper is subject to projection effects; second, to be maximally useful for subsequent cosmological analyses, $\lambdatrue$ should relate to the cluster mass as simply and directly as possible, with minimal intrinsic scatter. 

\subsubsection{\redmapper richness}
\redmapper defines a probability that a galaxy is a red cluster member above a threshold in $L/\lstar$, $\pmem$. This is summed over all possible cluster members to compute the richness of a given cluster:

\begin{equation}
\label{eqn:lambdaobs}
\begin{split}
\lambdaobs &\equiv  \sum_{\text{mem}} p(m \text{ is a red member} | \text{photometry}) \\
&= \sum_{\text{mem}} \pmem \\
&= \sum_{\text{mem}} p(\vec{x}|\lambda) \ p_{\text{free}} \ \theta_r \ \theta_i.
\end{split}
\end{equation}

In Equation \ref{eqn:lambdaobs}, the probability that a galaxy is a red member of a specified cluster is written as a product of four factors:

\begin{enumerate}
    \item $p(\vec{x}|\lambda)$ is the probability that a galaxy with observed properties $\vec{x}=(g-r, r-i, i-z, m_i, \text{R.A.}, \text{Dec.})$ (multiple photometric colors, \textit{i}-band magnitude, and position on the sky) is a red member of a  cluster of richness $\lambda$. This term is evaluated with a matched filter that is comprised of three sub-filters: the cluster galaxy radial number density profile, the cluster luminosity function, and the $\chi^2$ distribution comparing galaxy color with the red sequence model color at a given redshift;
    \item $p_{\text{free}}$ is the probability that the galaxy does not belong to another cluster
    \item $\theta_r$ is a radial weight function that acts as a smooth radial threshold for membership to account for the small photometric uncertainty on the position of a given candidate member;
    \item $\theta_i$ is a luminosity (\textit{i}-band magnitude) weight function that acts as a smooth luminosity threshold at $0.2 L_*$ for membership to account for the small photometric uncertainty on the apparent magnitude of a given candidate member.
\end{enumerate}

\subsubsection{Spectroscopic Richness}

We define our improved richness estimate, $\lambdatrue$, as:

\begin{equation}
\label{eqn:lambdaspec}
\begin{split}
\lambdatrue &\equiv  \sum_{\text{mem}} p(m \text{ is a red member} | \text{spectroscopy, photometry}) \\
&= \sum_{\text{mem}} \pspec \pred p_{\mathrm{free}} \theta_r \theta_i,\\
\end{split}
\end{equation}
 
where $\pspec$ is given by Equation \ref{eqn:halo_model} for members which are not the BCG and is set to unity for the BCG. This definition of $\lambdatrue$ achieves the primary goal of accounting for both spectroscopic information and galaxy color, and accounts for galaxy position and cluster percolation in the same way as \redmapper.

Notably, Eqn. \ref{eqn:lambdaspec} contains an independently computed $\pred$ whereas Eqn.  \ref{eqn:lambdaobs} does not. This $\pred$ developed by \citep{redmapper4} is computed from the $\chi_{\mathrm{s}}^2$ value representing the goodness of fit of the \redmapper red sequence template to the galaxy photometry, and modified to correct for photometric noise bias (for additional details, see \cite{redmapper4}):

\begin{equation}
    \pred (\chi_{\mathrm{s}}) = \frac{1}{2} \left[ 1 - \mathrm{erf} \left( \frac{\ln(\chi_{\mathrm{s}}/\chi_{\mathrm{ref}})}{\sqrt{2}\sigma} \right) \right],
\end{equation}

where $\chi_{\mathrm{ref}}$ and $\sigma$ are fit empirically by \cite{redmapper4} and found to be:
\begin{equation}
    \begin{split}
        \ln \chi_{\mathrm{ref}} = 2.44 \pm 0.08 \\
        \sigma = 0.28 \pm 0.11.
    \end{split}
\end{equation}

Combining $\pspec$ with $\pred$ unavoidably differs from the \redmapper-defined $\pmem$ because the matched filter that \redmapper uses to determine $p(\vec{x}|\lambda)$ contains sub-filters for the cluster density profile and the cluster luminosity function. It is necessary to construct $\lambdatrue$ in this way, rather than, for example, defining $\lambdatrue =\sum p_{\rm phot} \pspec$, because the \redmapper matched filter $p_{\rm phot}$ contains a galaxy background term that is correlated with the information provided by spectroscopy. For more information on the galaxy background term used to compute \redmapper richness, see \cite{redmapper1}.

Given this prescription for $\lambdatrue$, we estimate the richness bias due to projection effects by comparing $\lambdaobs$ with $\lambdatrue$ in each richness bin. For each richness bin $j$, we compute the richness bias from the candidate members $m$ passing our selection as follows:
\begin{align}
\label{eqn:projeffects}
    b_{\lambda,j} &\equiv \dfrac{\sum\limits_{m \in j}{p_{\rm phot}} - \sum\limits_{m \in j}{\pspec \pred p_{\mathrm{free}} \theta_r \theta_i}}{\sum\limits_{m \in j}{p_{\rm phot}}} \nonumber \\ 
    &= \dfrac{\sum \lambdaobs^{L\geq0.55\lstar} - \sum \lambdatrue^{L\geq0.55\lstar}}{\sum \lambdaobs^{L\geq0.55\lstar}}. 
\end{align}

As developed in \citet{redmapper4}, the mis-attribution of membership probability for galaxies along the line-of-sight can be dependent on photometric noise. Additionally, our measurement of $b_\lambda$ includes a contribution from the blue galaxy bias of the matched filter used to compute $p_{\rm phot}$.

\section{Results}
In this section we illustrate our model fit to the data (Section \ref{sec:modeling}), discuss the magnitude and richness dependence of observed projection effects (Section \ref{sec:mag_of_effect} \& \ref{sec:richness_dependence}), compare to other models from the literature (Section \ref{sec:sims}), and discuss the impact of our sample luminosity limit on our results (Section \ref{sec:sdss_brightness_test}).

\subsection{Fit of double-Gaussian model to data}
\label{sec:modeling}

\begin{figure*}
\includegraphics[scale=0.55]{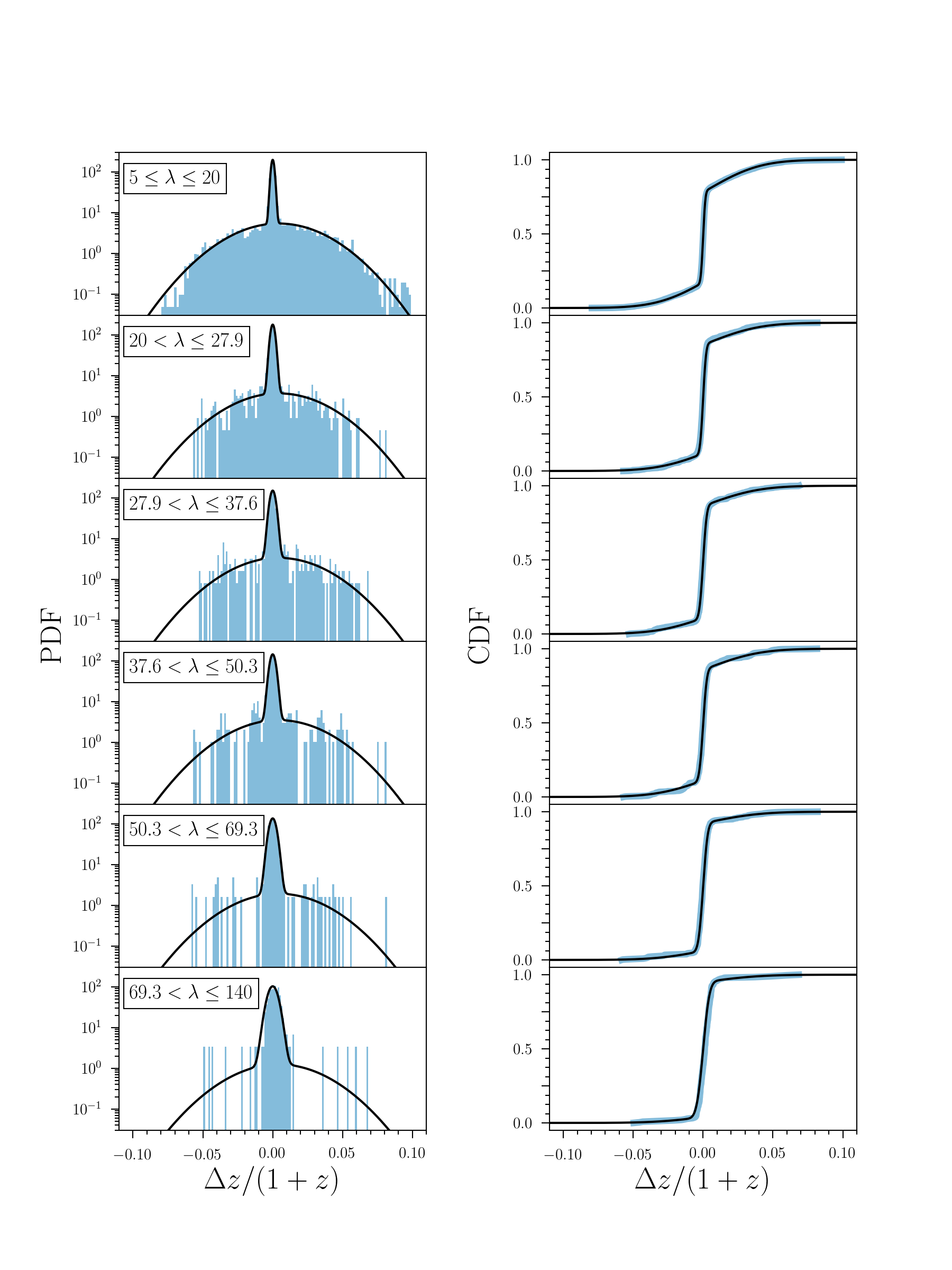}
\caption{\textit{Left}: Maximum likelihood model probability distribution functions (black) and histograms (blue) of the SDSS \redmapper cluster member velocities. The histograms have been normalized to unit area so that they are comparable to the model PDFs. \textit{Right}: Cumulative distribution functions of the data and model. We find a simple double-Gaussian model provides a good qualitative fit to the data.}\label{fig:sdss_maxL_model_and_data}
\end{figure*}

We first explore whether our double-Gaussian model can provide a reasonable description of projection effects in the low redshift SDSS \redmapper sample. Fig. \ref{fig:sdss_data} shows the distribution of line-of-sight velocities for candidate member galaxies, quantified in terms of relative difference in redshift to the BCG redshift, $\deltaznormed$, for the full ensemble of target clusters, spanning the redshift range $0.08 \leq z \leq 0.12$ and richness $\lambda \geq 5$, with the fiducial member galaxy luminosity cut, $L\geq0.55 \lstar$. The distribution shows two visually discernible velocity components that we can intuitively identify as cluster halo galaxies with virialized velocities (narrow component) and galaxies viewed in projection (broader component). The width of the broader velocity component is determined by the width of the \redmapper red sequence color model, which effectively serves as a photometric redshift cut. This two-component model for cluster membership is subject to the caveat that some galaxies that appear to be physically associated with the primary cluster halo due to their line-of-sight velocities may not in fact be inside the halo according to some other definition of halo membership, as discussed by \citet{Farahi2016}.

We find best fitting (maximum likelihood) double-Gaussian models for the spectroscopic data with the Nelder--Mead method \citep{neldermead}. The left panel of Fig. \ref{fig:sdss_maxL_model_and_data} shows these best fitting models overlaid on histograms of the spectroscopic data for the six independent bins of cluster richness. The right panel of Fig. \ref{fig:sdss_maxL_model_and_data} shows the cumulative distribution functions for the models and data. While the highest richness bin ($\lambda\geq69.3$) contains very few members, overall the model provides a good qualitative description of the data. The corresponding best fitting values are listed in Table \ref{tab:parameters}. We discuss additional validation tests in Appendix \ref{app:validation}.

\subsection{Magnitude of projection effects}
\label{sec:mag_of_effect}

\begin{figure*}
  \centering
  \includegraphics[scale=0.35]{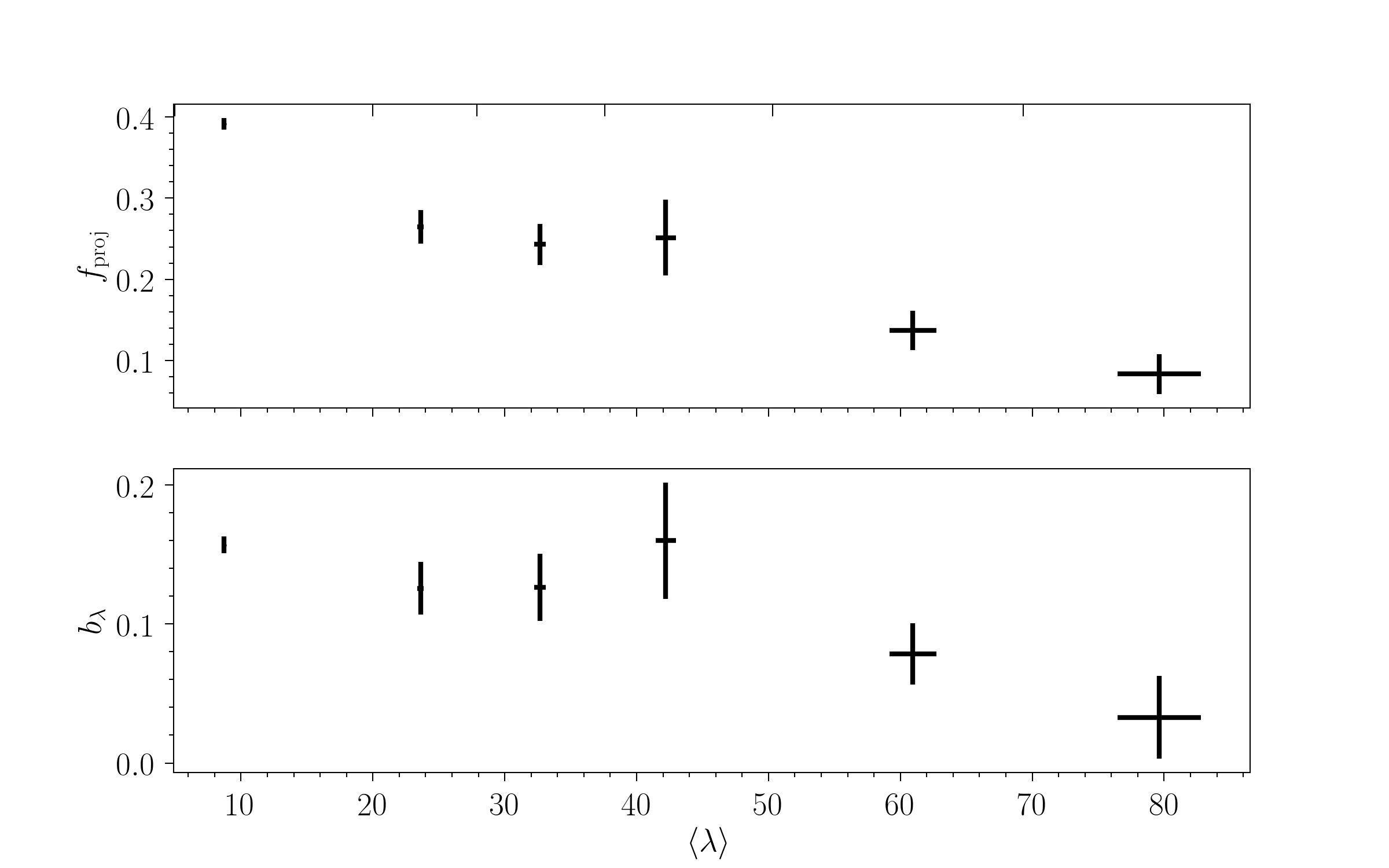}
  \caption{
    \textit{Top}: Model parameter $\fproj$ as a function of observed \redmapper richness, quantifying the probability that a candidate member galaxy identified by \redmapper algorithm is a line-of-sight projection. Upper $x$-axis tick marks indicate the edges of the richness bins used.  \textit{Bottom}: The richness bias $b_\lambda$ as a function of richness, quantifying the extent to which the richness \redmapper measures is biased due to projection effects. Both panels suggest a trend of increasing projection effects with decreasing richness (see Section~\ref{sec:richness_dependence} for details).
  }
  \label{fig:sdss_fproj_lambda_proj}
\end{figure*}

The posterior distribution for $\fproj$, quantifying the probability that a candidate member galaxy identified by the \redmapper algorithm in a given richness bin is a line-of-sight projection rather than a bona fide cluster member, is shown in the top panel of Fig.~\ref{fig:sdss_fproj_lambda_proj}. The result shows that galaxies in projection can account for as much as $40$ per cent (for $5\leq \lambda \leq 20$) of the candidate members in a given richness bin. 

The measurements of $\fproj$ shown in the top panel of Fig.~\ref{fig:sdss_fproj_lambda_proj} are independent of the astrophysical properties of the candidate cluster member galaxies, but dependent on the threshold of photometric membership probability for inclusion in the spectroscopic sample. $\fproj$ is a measure of the fraction of galaxies in the \redmapper catalogue that are not in fact associated with the clusters, but does not take into account the photometric weightings used to compute richness.

The \redmapper richness measure is designed to provide a {\it probabilistic} count of \emph{red} cluster members, weighting factors such as galaxy color and angular separation from the BCG. Following the formalism described in Section \ref{sec:formalism}, we can combine this information with our galaxy velocity measurements to compute a corrected mean spectroscopic richness, $\lambdatrue$, for each richness bin. The richness bias, defined by Equation \ref{eqn:projeffects}, is shown in the bottom panel of Fig.~\ref{fig:sdss_fproj_lambda_proj} for each richness bin. We see that as much as $10-20$ per cent of the richness in a given bin is associated with projection effects.

\subsection{Richness dependence of projection effects}
\label{sec:richness_dependence}
In addition to the overall amplitude, the most striking result revealed by Fig.~\ref{fig:sdss_fproj_lambda_proj} is the richness dependence of the measured projection effects. In terms of the fraction of galaxies that are viewed in projection (top panel) the results range from $f_{\rm{proj}} \sim 0.4$ for $5 \leq \lambda < 20$, to $f_{\rm{proj}} < 0.1$ for the highest richness bin. In terms of the richness bias estimates (bottom panel), the fractional corrections range from $\sim 16$ per cent in the lowest richness bin to $\sim 3$ per cent in highest richness bin. Fitting a linear model to the trend observed in $f_{\rm{proj}}$ with richness and in $b_\lambda$ with richness, we find that the statistical significance of a non-zero richness dependent trend is $>10\sigma$ and $>7\sigma$ confidence for $f_{\rm{proj}}$ and $b_\lambda$, respectively. We note that our measurement of projection effects in the $37.6 < \lambda \leq 50.3$ bin is relatively noisier than the other bins. The larger uncertainty due to sample variance of the measurements of $f_{\rm{proj}}$ and $b_\lambda$ in this bin is consistent with the explanation that its apparent deviation from monotonicity is likely a statistical fluctuation rather than some as-yet unmodeled systematic error. As discussed in \S \ref{sec:discussion}, for these low-redshift clusters, the uncertainty on this measurement will soon be greatly reduced with forthcoming data collected by the Dark Energy Spectroscopic Instrument \citep{desi}. These additional observations will enable us to verify this interpretation.

\subsection{Comparison with model predictions for projection effects}
\label{sec:sims}

\begin{figure*}
  \centering
  \includegraphics[scale=0.35]{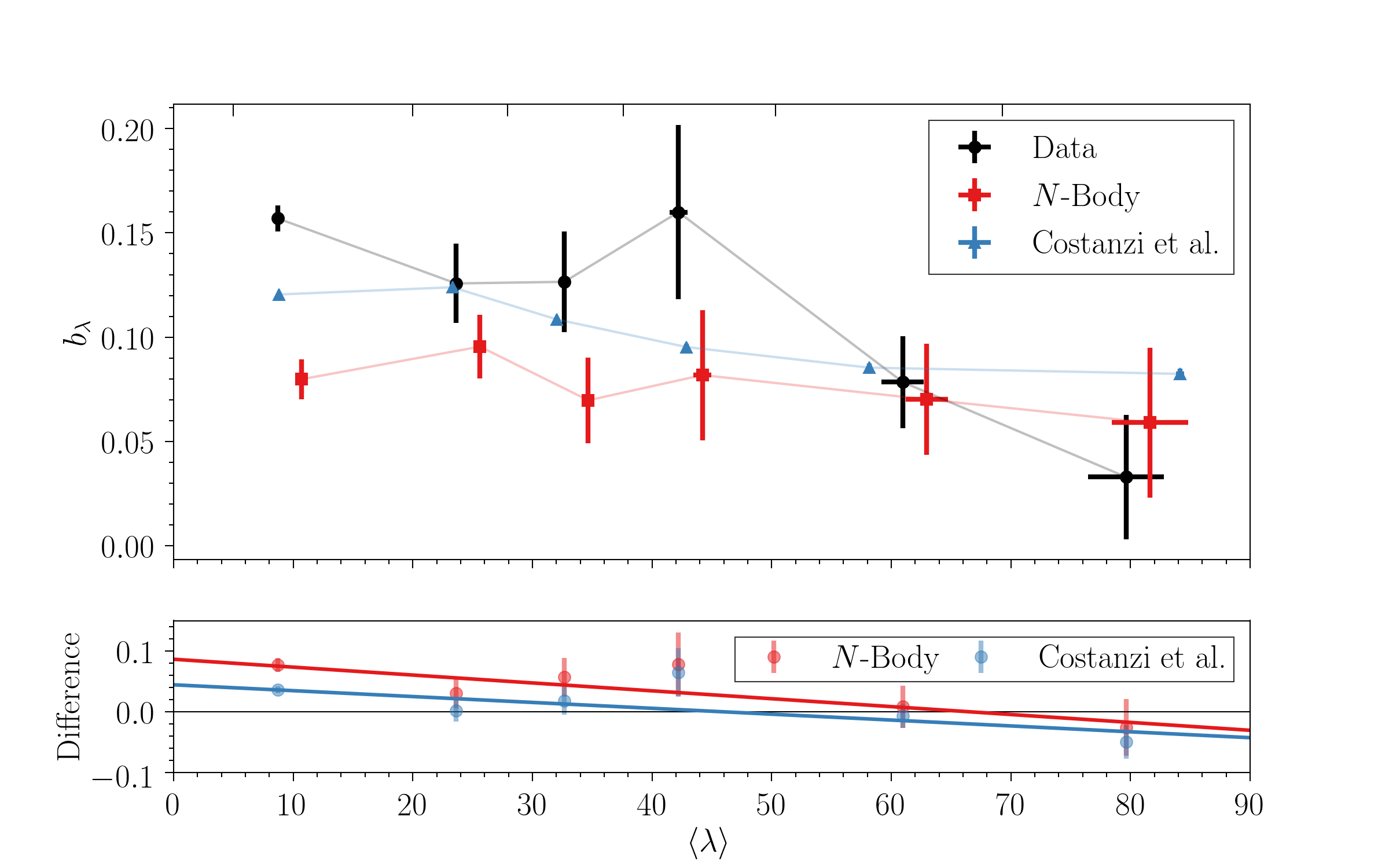}
  \caption{
    The richness bias, as a function of observed \redmapper richness, determined empirically from data (black), in galaxy cluster catalogues built from $N$-Body simulations \citep{derose2019} (red), and from the model of \citet{Costanzi1807.07072} (blue). Small offsets in the richness axis values of the $N$-Body curve have been introduced for clarity. Upper $x$-axis tick marks indicate the edges of the richness bins used. The data suggest a steeper empirical richness dependence of the spectroscopically measured projection effects than predicted by state-of-the-art models.
  }\label{fig:sdss_lambda_proj_data_v_sims}
\end{figure*}

We can compare our empirical measurement of projection effects with the predictions from two recent studies. In the first case, we compare our result to the model developed in \citet{Costanzi1807.07072} by combining data and simulation analyses. This projection effects model was employed in the recently published SDSS and DES Year 1 cluster cosmology analyses \citep[][respectively]{Costanzi2019,desy1}. Because the model is generated from a mock halo catalogue that does not contain individual galaxies, we cannot repeat our measurement on it directly.
Instead, we can compare our $\sigma_{\mathrm{proj}}$ and $b_\lambda$ measurements --- Equations \ref{eqn:likelihood} and \ref{eqn:projeffects}, respectively --- to the analogous redshift kernel for projections and $\frac{\lambda_\mathrm{obs} - \lambda_{\mathrm{true}}}{\lambda_{\mathrm{obs}}}$ of their model.
This comparison is important for two primary reasons: their model was tuned for the same SDSS catalogue as used in our work and was used in the DES Year 1 cluster cosmology analysis adopting the appropriate redshift kernel for projections calibrated on Y1 data \citep[see appendix A of][]{desy1}.

We also repeat our measurements on the mock \redmapper catalogue built from a catalogue of galaxies pasted onto dark matter haloes from the $N$-Body simulations underlying the \textsc{Buzzard} mock catalogues (\citealt{derose2019}; Wechsler et al. in preparation). Since the richness distributions in this catalogue differ somewhat from the data, we perform a simple abundance matching procedure to generate richness bins from the simulated catalogue. In brief, we scale the number of clusters in a given richness bin in the simulations by the ratio of the survey areas of our data and the simulations, and sequentially assign the highest richness clusters in \textsc{Buzzard} to successively lower richness bins. We can then repeat our measurement of line-of-sight velocities of the galaxies in the simulations in an identical manner to that performed on the data. 

We find consistency between our $\sigma_{\rm proj}$ measurement and the redshift kernel for projections estimated in \citet{Costanzi1807.07072} using \redmapper. Specifically, fitting a normal distribution to the stacked redshift kernel profiles derived in \citet{Costanzi1807.07072} we obtain $\sigma=0.0265$, comparable to our $\sigma_{\rm proj}=0.0290 \pm 0.0004$.
However, as shown in Fig. \ref{fig:sdss_lambda_proj_data_v_sims} we find that our empirical results exhibit a stronger richness dependence of the measured projection effects than either the \textsc{Buzzard} $N$-Body simulations or the model of \citet{Costanzi1807.07072}. While the \textsc{Buzzard} simulations predict a similar overall magnitude of projection effects to our data at the high richness end, they predict very little richness dependence of the signal. The results of \citet{Costanzi1807.07072} are comparable to our measurement for intermediate richness clusters, but show less dependence on richness than is evident in the data. In particular, a linear fit to the difference between the richness bias in data and the \textsc{Buzzard} simulation gives a slope inconsistent with zero at the $\sim 4\sigma$ level ($-0.0013 \pm 0.004$). The equivalent test comparing the data with the model from \citet{Costanzi1807.07072} yields a non-zero slope at the $\sim 3\sigma$ level ($-0.00097 \pm 0.0003$) (see lower panel of Fig. \ref{fig:sdss_lambda_proj_data_v_sims}). We interpret the combined results as evidence of steeper richness dependence in the data than in state-of-the-art models.

\subsection{Scaling of velocity dispersion with richness}
\label{sec:sigma_cl}

\begin{figure}
  \centering
  \includegraphics[scale=0.30]{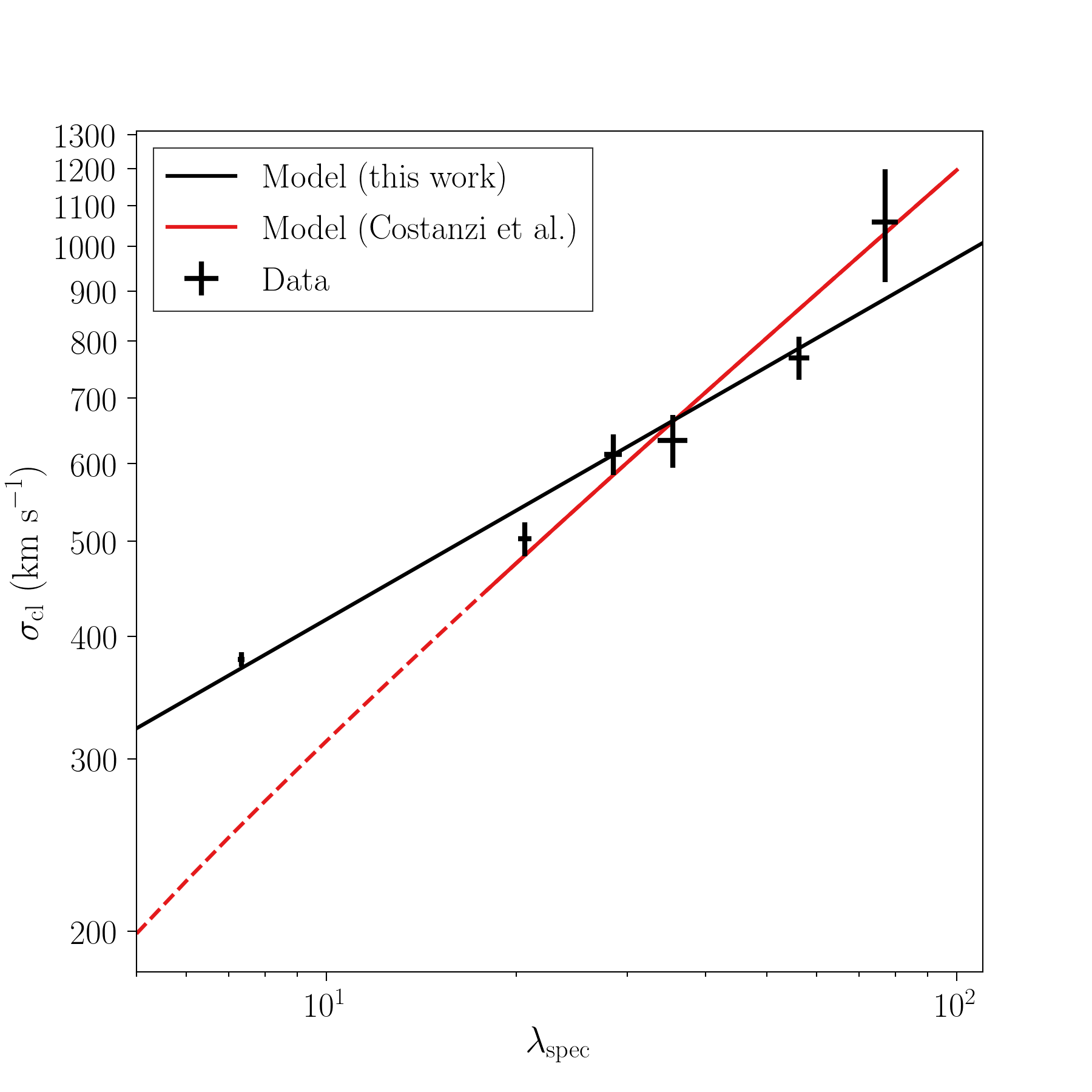}
  \caption{The observed mean cluster velocity dispersion, $\sigma_{\mathrm{cl}}$, as a function of corrected richness, $\lambda_{\mathrm{spec}}$. The best fitting power-law model has been overlaid. The best fitting slope of $0.37 \pm 0.04$ is comparable to that predicted from hydrodynamical simulations for the relationship between true (three-dimensional) galaxy velocity dispersion and halo mass, implying a power-law index for the $\lambdatrue$-halo mass relation of close to unity ($\alpha = 0.98\pm0.11$). For visual comparison we plot the relation implied by the best fitting mass-richness relation reported by \citep{Costanzi2019}, noting that the mass-richness relation from their work is calibrated with weak lensing mass data limited to clusters with photometric richness $\lambda \geq 20$ and a quantitative comparison with our work must propagate all uncertainties on model parameters.}
  \label{fig:sigma_cl}
\end{figure}

It is particularly interesting to compare the variation of $\sigma_{\rm cl}$ with the corrected richness, $\lambdatrue$. Fig. \ref{fig:sigma_cl} shows the relation, which is well described by a power-law model with index $0.37 \pm 0.04$ ($\chi_\nu^2$ = 2.0 for 4 degrees of freedom, corresponding to $p = 0.09$). Using hydrodynamical simulations incorporating physically motivated feedback schemes, \citet{Munari} predict a slope for the relation between galaxy velocity dispersion and halo mass for virialized systems, $\sigma_{\rm cl} \propto M^{0.364}$ (with a systematic uncertainty on the index of order $\sim 0.01$, depending on the precise feedback physics implementation). Together, these results imply an approximately linear relationship between $\lambdatrue$ and the three-dimensional halo mass, with $\lambdatrue$ scaling as mass to the power $\alpha = 0.98\pm0.11$. The implications of this result are discussed in Section~\ref{sec:discussion} below.

\subsection{Impact of the survey luminosity limit }
\label{sec:sdss_brightness_test}

\begin{figure*}
  \centering
  \includegraphics[scale=0.35]{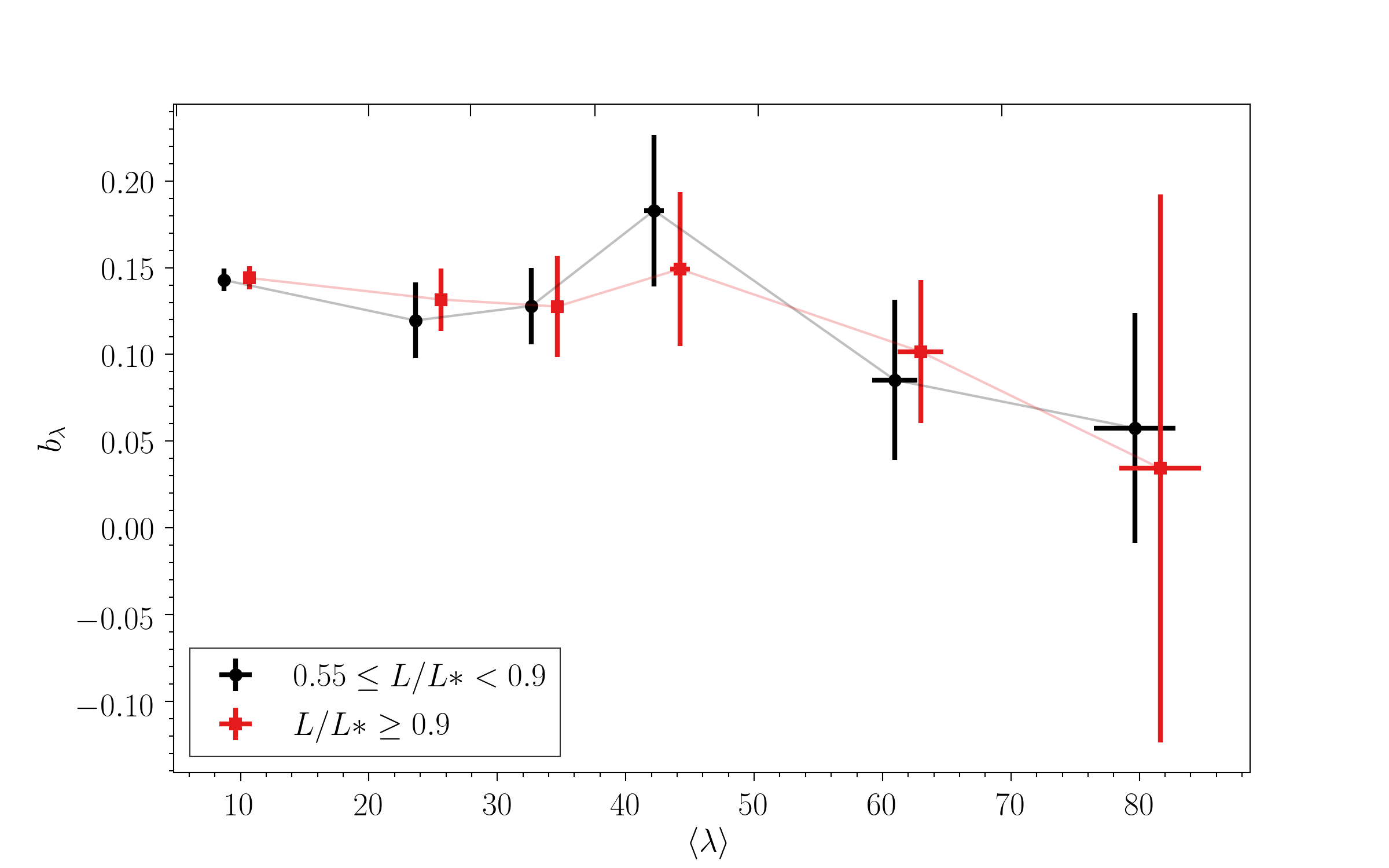}
  \caption{
    The richness bias, $b_\lambda$, as a function of observed \redmapper richness, for the two independent galaxy luminosity sub-samples with $0.55<L/\lstar<0.9$ and $L>0.9\lstar$. Small offsets in the richness axis values of the higher luminosity curve have been introduced for clarity. Upper $x$-axis tick marks indicate the edges of the richness bins used. We see no evidence for a galaxy luminosity dependence  of the projection effect results, suggesting the measured values may apply down to the $0.2 \lstar$ luminosity limit used by \redmapper.
  }\label{fig:sdss_lambda_proj_vs_richness_lum}
\end{figure*}

Our fiducial analysis is limited to galaxies with luminosities $L \geq 0.55 \lstar$, but, importantly, \redmapper includes galaxies as faint as $0.2L_{\star}$. In order to test for a possible galaxy luminosity dependence of the measured projection signal, we repeat the analysis of projection effects for two evenly sized luminosity bins corresponding to luminosities $0.55\leq L/\lstar<0.9$ and $L\geq0.9\lstar$, respectively. The results are shown in Fig. \ref{fig:sdss_lambda_proj_vs_richness_lum}. We see no evidence of a galaxy luminosity dependence of the measured projection effects over the luminosity range spanned by our data. While the SDSS spectroscopic sample is not complete to $0.2L_{\star}$ at any redshift spanned by the \redmapper sample, this result appears to rule out any strong dependence of the measurement on the luminosity limit of the survey. This dependence will be directly testable with the DESI spectroscopic sample.

\section{Discussion}
\label{sec:discussion}

Our study has  empirically quantified the incidence of projection effects on the SDSS DR8 \redmapper galaxy cluster catalogue. Utilizing complete spectral coverage for a well-defined subset of candidate cluster member galaxies available from SDSS at the low-redshift limit of the catalogue ($0.08 \leq z \leq 0.12$), we have demonstrated the ability to differentiate galaxies associated with the virialized cluster halo from galaxies viewed in projection. We have shown that a simple double-Gaussian model can be used to describe statistically the impact of projection effects. The observed incidence of projection effects in the SDSS \redmapper catalogue is substantial and exhibits a clear dependence on cluster richness, being several times stronger in typical low richness systems than the largest, richest clusters (Fig.~\ref{fig:sdss_fproj_lambda_proj}). The observed dependence of projection effects on richness is stronger than predicted by state-of-the-art models.

Our results are interesting in the context of the findings from two recent analyses that, respectively, utilized the SDSS and DES Year 1 \redmapper cluster catalogues to constrain cosmology \citep{Costanzi2019, desy1}. In particular, the DES Year 1 analysis yielded surprisingly low values for both $S_8 \equiv \sigma_8 (\Omega_m/0.3)^{0.5} = 0.65 \pm 0.04$, and the power-law index of the scaling between the `true' (three-dimensional) richness and halo mass, $\alpha=0.745\pm0.045$. The $\sigma_8-\Omega_m$ posteriors reported from that analysis are in $2.4 \sigma$ tension with the DES Year 1 cosmological constraints from galaxy clustering and weak lensing \citep[$3\times2$-point;][]{desy1_3x2pt}, and in $5.6 \sigma$ tension with the \textit{Planck} CMB analysis \citep{Planck2018a}.

\citet{desy1} noted that restricting their analysis to a higher richness threshold ($\lambda \geq 30$) reduced the tension with other probes, pointing to one or more richness-dependent effects as the likely source of the tension. Both the SDSS and DES cluster analyses used the prescription of \citet{Costanzi1807.07072} to model projection effects. Our study has shown that there are significant, richness-dependent effects not accounted for by that model (Section \ref{sec:sims}). Our finding of a steeper inverse dependence of $b_\lambda$ on observed richness suggests a steeper intrinsic slope of weak lensing mass with true richness than reported by \citet{desy1}. 
Moreover, the slope of the true richness--mass relation implied by our $\lambda_{\mathrm{spec}}$ and $\sigma_{\mathrm{cl}}$ measurements, $\alpha=0.98\pm0.11$, is significantly steeper than the \citet{desy1} value, and consistent with theoretical predictions (Section~\ref{sec:sigma_cl}).

Our results also have implications for the interpretation of weak lensing mass measurements for \redmapper clusters. Fig. \ref{fig:sdss_fproj_lambda_proj} shows that, for low-to-intermediate richness systems, a significant fraction of the galaxies identified by \redmapper as being associated with a cluster will typically be line-of-sight projections. The mass associated with these projected galaxies will also boost the measured lensing signals. However, the lower mass-to-light ratios for field galaxies compared to cluster haloes ($5-10\times$; \citealt{Dai0911.2230}) will lead to these lensing boosts being modest ($\sim$ few per cent). The net result is that, for the richest clusters, both richness and weak lensing mass should be measured relatively accurately while, for the least rich systems, projection effects will cause richness to be biased high and the mass at a given richness to be biased low. 

Future work will explore quantitatively the impact of our new, empirical calibration of projection effects on the cosmological constraints from the SDSS and DES \redmapper cluster catalogues. While firm conclusions must await these full analyses, we note that the degeneracy between $\Omega_m$ and the true richness-mass relation slope reported by \citet{desy1} suggests that a value for $\alpha$ closer to unity would shift the inferred $\Omega_m$ constraints toward a concordance cosmology. This is also consistent with the recent results of \cite{to2020, to2020a}, who combined DES weak lensing and galaxy clustering data with cluster clustering, cluster abundances, and cluster-galaxy cross-correlations \footnote{The analyses of \cite{to2020, to2020a} remove essentially all information on the mass calibration from length scales $< 10$ Mpc, reducing the sensitivity to the projection effects discussed here.} finding no significant tension with the results from \cite{desy1}.

Our work points to the potential utility of $\lambdatrue$ as a new, lower scatter mass proxy for cosmological studies. Previous work (e.g. \citealt{Mantz2010a, Mantz2015, wu2010}) has demonstrated how the availability of such mass proxy measurements for even a modest fraction of the clusters in a survey can substantially improve the cosmological constraining power of that survey, and provide unique insights into the form, scatter, and evolution of the key mass-observable scaling relations (see also \citealt{Allen1103.4829}). Utilizing $\lambdatrue$ measurements to their fullest extent will require accumulating extensive spectroscopic observations for individual clusters. In combination with X-ray observations, such measurements hold the potential to provide powerful complementary constraints on halo properties. We emphasize that our modelling approach also provides robust, stacked velocity dispersion estimates for clusters in selected richness bins, which brings an additional route to constrain cluster masses directly. Future work will examine the utility of such measurements, as a complement to weak lensing data, in extracting cosmological constraints. In addition to investigating the use of stacked spectroscopic cluster velocity dispersions as a mass proxy for cluster cosmology experiments, we leave it to future work to investigate any relationship between the weak lensing signals for subsamples of clusters binned on spectroscopic richness or the spectroscopically estimated extent of projection effects.

Our measurements suggest that relatively modest observational campaigns targeting a representative subsample of clusters is sufficient to adequately calibrate the projection effects model and thus make use of the large optical cluster catalogues from surveys such as the Dark Energy Survey and the Rubin Observatory Legacy Survey of Space and Time. While our work excludes a strong galaxy luminosity dependence of the measured projection signal, future work should extend the spectroscopic follow-up of member galaxies to the low luminosity limit employed by the \redmapper algorithm, $L>0.2\lstar$. For low-redshift clusters ($z<0.2)$, this will soon be possible using the Dark Energy Spectroscopic Instrument \citep{desi}. We note additionally the publicly available HeCS-red spectroscopic survey of \redmapper cluster members \citep{Rines2018}. This survey combines archival data from \citet{Rines2013, Rines2016} with new spectra for a sample of 27 clusters of the 30 \redmapper clusters satisfying the selection function $0.1 \leq z \leq 0.25$, $\lambda \geq 64$, declination greater than 10 degrees, and right ascension less than 9h or greater than 20h. The stated limiting magnitude of $r = 20.0$ from \citet{Rines2018} at the low-redshift end of this sample implies that this sample is complete down to $0.2\lstar$ at redshift $z = 0.12$. We thus conclude that an extended version of our work should be conducted to include members satisfying $L\geq0.55\lstar$ from the clusters with $z_\lambda \leq 0.12$ in this survey. The additional members at this high richness end are expected to reduce the uncertainty of our measurements. These observations could, in conjunction with additional follow-up spectroscopy of lower richness SDSS \redmapper clusters (see e.g. \citealt{Sohn2010.05817}), be used to extend our analysis down to $0.2 \lstar$ for all richness bins. For higher redshift systems, this will require dedicated follow-up spectroscopy with large aperture telescopes. Programs to gather these data with the Gemini Observatory, the Magellan Telescopes at Carnegie Observatories, and the W.M. Keck Observatory have begun. Further, natural extensions of the work would explore the cluster-to-cluster variation and the radial dependence of projection effects with respect to cluster centers (see also \citealt{Tomooka2020}). In addition to the radial dependence of projection effects, spatial distribution analysis may prove useful for determining the average physical size of clusters  \citep{Tomooka2020, Aung2020, Wagoner2020} which, together with velocity dispersions, may provide additional mass constraints. Finally, a limitation of the selection applied to our spectroscopic sample is potential contamination due to mis-centered clusters, for which the nominal BCG is not the correct reference from which to compute $\deltaznormed$. We note, however, that most nominal BCGs will still be cluster members, causing any bias in the line-of-sight velocities to be less than the cluster velocity dispersion. We defer a more sophisticated treatment of the cluster BCGs to future analyses for which we expect to have spectroscopic redshift coverage for nearly all members of individual clusters.


\section{Conclusions}
\label{sec:conclusion}
We have investigated the extent to which projection effects impact the Sloan Digital Sky Survey (SDSS) \redmapper galaxy cluster catalogue. Using the complete spectral coverage available for luminous galaxies at the low redshift limit ($0.08\leq z \leq0.12$) of the catalogue, we have quantified the influence of projection effects on the measured cluster richness. Our findings can be summarized as follows:
 
\begin{enumerate}
\item A simple double-Gaussian model, with one Gaussian describing the virialized cluster galaxies and the other the projected component, is sufficient to describe the impact of projection effects in the SDSS \redmapper sample. 
\item The observed incidence of projection effects is substantial, with $\sim 40$ per cent of candidate member galaxies and $\sim 16$ per cent of the richness being associated with projection effects for objects with apparent richness in the range $5\leq \lambda < 20$.
\item The amplitude of projection effects in the SDSS \redmapper catalogue is a function of richness, with projection effects being greater in the lowest richness objects.  
\item The observed amplitude of projection effects is comparable to that of state-of-the-art models, but has steeper richness dependence than these models. 
\item The slope of the velocity dispersion--$\lambdatrue$ relation implies an approximately linear relationship between $\lambdatrue$ and the three-dimensional halo mass.
\item Splitting the member galaxy catalogue into low and high luminosity sub-samples, no clear galaxy luminosity dependence of the observed projection effect signal is observed.
\end{enumerate} 
 
 Our results provide a robust, empirical description of the impact of projection effects in the low redshift limit of the SDSS \redmapper cluster catalogue. The $\lambdatrue$ mass proxy, combining photometric and spectroscopic information, and associated robust measurements of cluster velocity dispersions in richness bins, opens the possibility to extract improved constraints on astrophysics and cosmology from the SDSS \redmapper and other cluster samples. Our work exemplifies the essential synergies between optical imaging and spectroscopic studies of galaxy clusters. Future work, also using \textit{Chandra} follow-up observations for a complete sub-sample of the clusters studied here, will quantify the improvements in our knowledge of the mass--richness scaling relation obtained with $\lambdatrue$ over $\lambdaobs$ measurements.
 Work to gather $\lambdatrue$ measurements for clusters spanning the full redshift and galaxy luminosity range of the SDSS and DES galaxy cluster catalogues is underway. 

\section*{Acknowledgements}

The authors would like to thank Gus Evrard and Arya Farahi for helpful discussion.

This work was supported by the U.S. Department of Energy under contract number DE-AC02-76SF00515 and by \textit{Chandra}  Award  Number  GO8-19101A, issued  by  the  \textit{Chandra} X-ray Observatory Center.

JM thanks the LSSTC Data Science Fellowship Program, which is funded by LSSTC, NSF Cybertraining Grant \#1829740, the Brinson Foundation, and the Moore Foundation; his participation in the program has benefited this work. This work was supported by the Department of Energy, Laboratory Directed Research and Development program at SLAC National Accelerator Laboratory, under contract DE-AC02-76SF00515 and as part of the Panofsky Fellowship awarded to DG. MC is supported by the ERC-StG 'ClustersXCosmo' grant agreement 716762. TJ acknowledges support from the U.S. Department of Energy, Office of Science, Office of High Energy Physics, under Award Number A00-1465-001.

Funding for the Sloan Digital Sky Survey IV has been provided by the Alfred P. Sloan Foundation, the U.S. Department of Energy Office of Science, and the Participating Institutions. SDSS-IV acknowledges
support and resources from the Center for High-Performance Computing at
the University of Utah. The SDSS web site is www.sdss.org.

SDSS-IV is managed by the Astrophysical Research Consortium for the 
Participating Institutions of the SDSS Collaboration including the 
Brazilian Participation Group, the Carnegie Institution for Science, 
Carnegie Mellon University, the Chilean Participation Group, the French Participation Group, Harvard-Smithsonian Center for Astrophysics, 
Instituto de Astrof\'isica de Canarias, The Johns Hopkins University, Kavli Institute for the Physics and Mathematics of the Universe (IPMU) / 
University of Tokyo, the Korean Participation Group, Lawrence Berkeley National Laboratory, 
Leibniz Institut f\"ur Astrophysik Potsdam (AIP),  
Max-Planck-Institut f\"ur Astronomie (MPIA Heidelberg), 
Max-Planck-Institut f\"ur Astrophysik (MPA Garching), 
Max-Planck-Institut f\"ur Extraterrestrische Physik (MPE), 
National Astronomical Observatories of China, New Mexico State University, 
New York University, University of Notre Dame, 
Observat\'ario Nacional / MCTI, The Ohio State University, 
Pennsylvania State University, Shanghai Astronomical Observatory, 
United Kingdom Participation Group,
Universidad Nacional Aut\'onoma de M\'exico, University of Arizona, 
University of Colorado Boulder, University of Oxford, University of Portsmouth, 
University of Utah, University of Virginia, University of Washington, University of Wisconsin, 
Vanderbilt University, and Yale University.

\section*{Data availability}
The data underlying this article are available in the SDSS Data Release 8 repository, at \url{http://www.sdss3.org/dr8/}.

Publicly released \redmapper catalogues constructed from these data are available at \url{http://risa.stanford.edu/redmapper/}.



\bibliography{sdss-proj-effects} 

\begin{thebibliography}{}
\makeatletter
\relax
\def\mn@urlcharsother{\let\do\@makeother \do\$\do\&\do\#\do\^\do\_\do\%\do\~}
\def\mn@doi{\begingroup\mn@urlcharsother \@ifnextchar [ {\mn@doi@}
  {\mn@doi@[]}}
\def\mn@doi@[#1]#2{\def\@tempa{#1}\ifx\@tempa\@empty \href
  {http://dx.doi.org/#2} {doi:#2}\else \href {http://dx.doi.org/#2} {#1}\fi
  \endgroup}
\def\mn@eprint#1#2{\mn@eprint@#1:#2::\@nil}
\def\mn@eprint@arXiv#1{\href {http://arxiv.org/abs/#1} {{\tt arXiv:#1}}}
\def\mn@eprint@dblp#1{\href {http://dblp.uni-trier.de/rec/bibtex/#1.xml}
  {dblp:#1}}
\def\mn@eprint@#1:#2:#3:#4\@nil{\def\@tempa {#1}\def\@tempb {#2}\def\@tempc
  {#3}\ifx \@tempc \@empty \let \@tempc \@tempb \let \@tempb \@tempa \fi \ifx
  \@tempb \@empty \def\@tempb {arXiv}\fi \@ifundefined
  {mn@eprint@\@tempb}{\@tempb:\@tempc}{\expandafter \expandafter \csname
  mn@eprint@\@tempb\endcsname \expandafter{\@tempc}}}

\bibitem[\protect\citeauthoryear{{Abbott} et~al.,}{{Abbott}
  et~al.}{2018}]{desy1_3x2pt}
{Abbott} T.~M.~C.,  et~al., 2018, \mn@doi [\prd] {10.1103/PhysRevD.98.043526},
  \href {https://ui.adsabs.harvard.edu/abs/2018PhRvD..98d3526A} {98, 043526}

\bibitem[\protect\citeauthoryear{{Abbott} et~al.,}{{Abbott}
  et~al.}{2020}]{desy1}
{Abbott} T.~M.~C.,  et~al., 2020, \mn@doi [\prd] {10.1103/PhysRevD.102.023509},
  \href {https://ui.adsabs.harvard.edu/abs/2020PhRvD.102b3509A} {102, 023509}

\bibitem[\protect\citeauthoryear{{Ahn} et~al.,}{{Ahn} et~al.}{2014}]{sdss_spec}
{Ahn} C.~P.,  et~al., 2014, \mn@doi [\apjs] {10.1088/0067-0049/211/2/17}, \href
  {http://adsabs.harvard.edu/abs/2014ApJS..211...17A} {211, 17}

\bibitem[\protect\citeauthoryear{{Aihara} et~al.,}{{Aihara}
  et~al.}{2011}]{sdss_dr8}
{Aihara} H.,  et~al., 2011, \mn@doi [\apjs] {10.1088/0067-0049/193/2/29}, \href
  {http://adsabs.harvard.edu/abs/2011ApJS..193...29A} {193, 29}

\bibitem[\protect\citeauthoryear{{Allen}, {Evrard}  \& {Mantz}}{{Allen}
  et~al.}{2011}]{Allen1103.4829}
{Allen} S.~W.,  {Evrard} A.~E.,   {Mantz} A.~B.,  2011, \mn@doi [\araa]
  {10.1146/annurev-astro-081710-102514}, \href
  {https://ui.adsabs.harvard.edu/abs/2011ARA&A..49..409A} {49, 409}

\bibitem[\protect\citeauthoryear{{Applegate} et~al.,}{{Applegate}
  et~al.}{2014}]{WTGIII}
{Applegate} D.~E.,  et~al., 2014, \mn@doi [\mnras] {10.1093/mnras/stt2129},
  \href {https://ui.adsabs.harvard.edu/abs/2014MNRAS.439...48A} {439, 48}

\bibitem[\protect\citeauthoryear{{Aung}, {Nagai}, {Rozo}  \& {Garcia}}{{Aung}
  et~al.}{2020}]{Aung2020}
{Aung} H.,  {Nagai} D.,  {Rozo} E.,   {Garcia} R.,  2020, arXiv e-prints, \href
  {https://ui.adsabs.harvard.edu/abs/2020arXiv200311557A} {p. arXiv:2003.11557}

\bibitem[\protect\citeauthoryear{Benson et~al.,}{Benson
  et~al.}{2013}]{Benson_2013}
Benson B.~A.,  et~al., 2013, \mn@doi [The Astrophysical Journal]
  {10.1088/0004-637x/763/2/147}, 763, 147

\bibitem[\protect\citeauthoryear{{Bocquet} et~al.,}{{Bocquet}
  et~al.}{2019}]{Bocquet2019}
{Bocquet} S.,  et~al., 2019, \mn@doi [\apj] {10.3847/1538-4357/ab1f10}, \href
  {https://ui.adsabs.harvard.edu/abs/2019ApJ...878...55B} {878, 55}

\bibitem[\protect\citeauthoryear{{Cataneo} \& {Rapetti}}{{Cataneo} \&
  {Rapetti}}{2018}]{Cataneo2018b}
{Cataneo} M.,  {Rapetti} D.,  2018, \mn@doi [International Journal of Modern
  Physics D] {10.1142/S0218271818480061}, \href
  {https://ui.adsabs.harvard.edu/abs/2018IJMPD..2748006C} {27, 1848006}

\bibitem[\protect\citeauthoryear{{Costanzi} et~al.,}{{Costanzi}
  et~al.}{2019a}]{Costanzi1807.07072}
{Costanzi} M.,  et~al., 2019a, \mn@doi [\mnras] {10.1093/mnras/sty2665}, \href
  {https://ui.adsabs.harvard.edu/abs/2019MNRAS.482..490C} {482, 490}

\bibitem[\protect\citeauthoryear{{Costanzi} et~al.,}{{Costanzi}
  et~al.}{2019b}]{Costanzi2019}
{Costanzi} M.,  et~al., 2019b, \mn@doi [\mnras] {10.1093/mnras/stz1949}, \href
  {https://ui.adsabs.harvard.edu/abs/2019MNRAS.488.4779C} {488, 4779}

\bibitem[\protect\citeauthoryear{{DESI Collaboration} et~al.,}{{DESI
  Collaboration} et~al.}{2016}]{desi}
{DESI Collaboration} et~al., 2016, arXiv e-prints, \href
  {https://ui.adsabs.harvard.edu/abs/2016arXiv161100036D} {p. arXiv:1611.00036}

\bibitem[\protect\citeauthoryear{{Dai}, {Bregman}, {Kochanek}  \&
  {Rasia}}{{Dai} et~al.}{2010}]{Dai0911.2230}
{Dai} X.,  {Bregman} J.~N.,  {Kochanek} C.~S.,   {Rasia} E.,  2010, \mn@doi
  [\apj] {10.1088/0004-637X/719/1/119}, \href
  {https://ui.adsabs.harvard.edu/abs/2010ApJ...719..119D} {719, 119}

\bibitem[\protect\citeauthoryear{{DeRose} et~al.,}{{DeRose}
  et~al.}{2019}]{derose2019}
{DeRose} J.,  et~al., 2019, arXiv e-prints, \href
  {https://ui.adsabs.harvard.edu/abs/2019arXiv190102401D} {p. arXiv:1901.02401}

\bibitem[\protect\citeauthoryear{{Dietrich} et~al.,}{{Dietrich}
  et~al.}{2019}]{Dietrich2019}
{Dietrich} J.~P.,  et~al., 2019, \mn@doi [\mnras] {10.1093/mnras/sty3088},
  \href {https://ui.adsabs.harvard.edu/abs/2019MNRAS.483.2871D} {483, 2871}

\bibitem[\protect\citeauthoryear{{Eckert}, {Finoguenov}, {Ghirardini},
  {Grandis}, {Kaefer}, {Sanders}  \& {Ramos-Ceja}}{{Eckert}
  et~al.}{2020}]{erosita1}
{Eckert} D.,  {Finoguenov} A.,  {Ghirardini} V.,  {Grandis} S.,  {Kaefer} F.,
  {Sanders} J.,   {Ramos-Ceja} M.,  2020, \mn@doi [The Open Journal of
  Astrophysics] {10.21105/astro.2009.03944}, \href
  {https://ui.adsabs.harvard.edu/abs/2020OJAp....3E..12E} {3, 12}

\bibitem[\protect\citeauthoryear{{Eisenstein} et~al.,}{{Eisenstein}
  et~al.}{2011}]{sdss3}
{Eisenstein} D.~J.,  et~al., 2011, \mn@doi [\aj] {10.1088/0004-6256/142/3/72},
  \href {https://ui.adsabs.harvard.edu/abs/2011AJ....142...72E} {142, 72}

\bibitem[\protect\citeauthoryear{{Farahi}, {Evrard}, {Rozo}, {Rykoff}  \&
  {Wechsler}}{{Farahi} et~al.}{2016}]{Farahi2016}
{Farahi} A.,  {Evrard} A.~E.,  {Rozo} E.,  {Rykoff} E.~S.,   {Wechsler} R.~H.,
  2016, \mn@doi [\mnras] {10.1093/mnras/stw1143}, \href
  {https://ui.adsabs.harvard.edu/abs/2016MNRAS.460.3900F} {460, 3900}

\bibitem[\protect\citeauthoryear{{Heneka}, {Rapetti}, {Cataneo}, {Mantz},
  {Allen}  \& {von der Linden}}{{Heneka} et~al.}{2018}]{Cataneo2018a}
{Heneka} C.,  {Rapetti} D.,  {Cataneo} M.,  {Mantz} A.~B.,  {Allen} S.~W.,
  {von der Linden} A.,  2018, \mn@doi [\mnras] {10.1093/mnras/stx2549}, \href
  {https://ui.adsabs.harvard.edu/abs/2018MNRAS.473.3882H} {473, 3882}

\bibitem[\protect\citeauthoryear{{Hoekstra}, {Herbonnet}, {Muzzin}, {Babul},
  {Mahdavi}, {Viola}  \& {Cacciato}}{{Hoekstra} et~al.}{2015}]{Hoekstra2015}
{Hoekstra} H.,  {Herbonnet} R.,  {Muzzin} A.,  {Babul} A.,  {Mahdavi} A.,
  {Viola} M.,   {Cacciato} M.,  2015, \mn@doi [\mnras] {10.1093/mnras/stv275},
  \href {https://ui.adsabs.harvard.edu/abs/2015MNRAS.449..685H} {449, 685}

\bibitem[\protect\citeauthoryear{{Mana}, {Giannantonio}, {Weller}, {Hoyle},
  {H{\"u}tsi}  \& {Sartoris}}{{Mana} et~al.}{2013}]{mana2013}
{Mana} A.,  {Giannantonio} T.,  {Weller} J.,  {Hoyle} B.,  {H{\"u}tsi} G.,
  {Sartoris} B.,  2013, \mn@doi [\mnras] {10.1093/mnras/stt1062}, \href
  {https://ui.adsabs.harvard.edu/abs/2013MNRAS.434..684M} {434, 684}

\bibitem[\protect\citeauthoryear{{Mantz}, {Allen}, {Rapetti}  \&
  {Ebeling}}{{Mantz} et~al.}{2010a}]{Mantz2010a}
{Mantz} A.,  {Allen} S.~W.,  {Rapetti} D.,   {Ebeling} H.,  2010a, \mn@doi
  [\mnras] {10.1111/j.1365-2966.2010.16992.x}, \href
  {https://ui.adsabs.harvard.edu/abs/2010MNRAS.406.1759M} {406, 1759}

\bibitem[\protect\citeauthoryear{{Mantz}, {Allen}, {Ebeling}, {Rapetti}  \&
  {Drlica-Wagner}}{{Mantz} et~al.}{2010b}]{Mantz2010b}
{Mantz} A.,  {Allen} S.~W.,  {Ebeling} H.,  {Rapetti} D.,   {Drlica-Wagner} A.,
   2010b, \mn@doi [\mnras] {10.1111/j.1365-2966.2010.16993.x}, \href
  {https://ui.adsabs.harvard.edu/abs/2010MNRAS.406.1773M} {406, 1773}

\bibitem[\protect\citeauthoryear{{Mantz}, {Allen}  \& {Rapetti}}{{Mantz}
  et~al.}{2010c}]{Mantz2010c}
{Mantz} A.,  {Allen} S.~W.,   {Rapetti} D.,  2010c, \mn@doi [\mnras]
  {10.1111/j.1365-2966.2010.16794.x}, \href
  {https://ui.adsabs.harvard.edu/abs/2010MNRAS.406.1805M} {406, 1805}

\bibitem[\protect\citeauthoryear{{Mantz} et~al.,}{{Mantz}
  et~al.}{2015}]{Mantz2015}
{Mantz} A.~B.,  et~al., 2015, \mn@doi [\mnras] {10.1093/mnras/stu2096}, \href
  {https://ui.adsabs.harvard.edu/abs/2015MNRAS.446.2205M} {446, 2205}

\bibitem[\protect\citeauthoryear{{McClintock} et~al.,}{{McClintock}
  et~al.}{2019}]{McClintock2019}
{McClintock} T.,  et~al., 2019, \mn@doi [\mnras] {10.1093/mnras/sty2711}, \href
  {https://ui.adsabs.harvard.edu/abs/2019MNRAS.482.1352M} {482, 1352}

\bibitem[\protect\citeauthoryear{{Munari}, {Biviano}, {Borgani}, {Murante}  \&
  {Fabjan}}{{Munari} et~al.}{2013}]{Munari}
{Munari} E.,  {Biviano} A.,  {Borgani} S.,  {Murante} G.,   {Fabjan} D.,  2013,
  \mn@doi [\mnras] {10.1093/mnras/stt049}, \href
  {https://ui.adsabs.harvard.edu/abs/2013MNRAS.430.2638M} {430, 2638}

\bibitem[\protect\citeauthoryear{Nelder \& Mead}{Nelder \&
  Mead}{1965}]{neldermead}
Nelder J.~A.,  Mead R.,  1965, \mn@doi [The Computer Journal]
  {10.1093/comjnl/7.4.308}, 7, 308

\bibitem[\protect\citeauthoryear{{Planck Collaboration} et~al.,}{{Planck
  Collaboration} et~al.}{2018}]{Planck2018a}
{Planck Collaboration} et~al., 2018, arXiv e-prints, \href
  {https://ui.adsabs.harvard.edu/abs/2018arXiv180706209P} {p. arXiv:1807.06209}

\bibitem[\protect\citeauthoryear{{Reiprich} \& {B{\"o}hringer}}{{Reiprich} \&
  {B{\"o}hringer}}{2002}]{reiprich2002}
{Reiprich} T.~H.,  {B{\"o}hringer} H.,  2002, \mn@doi [\apj] {10.1086/338753},
  \href {https://ui.adsabs.harvard.edu/abs/2002ApJ...567..716R} {567, 716}

\bibitem[\protect\citeauthoryear{{Rines}, {Geller}, {Diaferio}  \&
  {Kurtz}}{{Rines} et~al.}{2013}]{Rines2013}
{Rines} K.,  {Geller} M.~J.,  {Diaferio} A.,   {Kurtz} M.~J.,  2013, \mn@doi
  [\apj] {10.1088/0004-637X/767/1/15}, \href
  {https://ui.adsabs.harvard.edu/abs/2013ApJ...767...15R} {767, 15}

\bibitem[\protect\citeauthoryear{{Rines}, {Geller}, {Diaferio}  \&
  {Hwang}}{{Rines} et~al.}{2016}]{Rines2016}
{Rines} K.~J.,  {Geller} M.~J.,  {Diaferio} A.,   {Hwang} H.~S.,  2016, \mn@doi
  [\apj] {10.3847/0004-637X/819/1/63}, \href
  {https://ui.adsabs.harvard.edu/abs/2016ApJ...819...63R} {819, 63}

\bibitem[\protect\citeauthoryear{{Rines}, {Geller}, {Diaferio}, {Hwang}  \&
  {Sohn}}{{Rines} et~al.}{2018}]{Rines2018}
{Rines} K.~J.,  {Geller} M.~J.,  {Diaferio} A.,  {Hwang} H.~S.,   {Sohn} J.,
  2018, \mn@doi [\apj] {10.3847/1538-4357/aacd49}, \href
  {https://ui.adsabs.harvard.edu/abs/2018ApJ...862..172R} {862, 172}

\bibitem[\protect\citeauthoryear{{Rozo} \& {Rykoff}}{{Rozo} \&
  {Rykoff}}{2014}]{redmapper2}
{Rozo} E.,  {Rykoff} E.~S.,  2014, \mn@doi [\apj] {10.1088/0004-637X/783/2/80},
  \href {https://ui.adsabs.harvard.edu/abs/2014ApJ...783...80R} {783, 80}

\bibitem[\protect\citeauthoryear{{Rozo}, {Rykoff}, {Bartlett}  \&
  {Melin}}{{Rozo} et~al.}{2015a}]{redmapper3}
{Rozo} E.,  {Rykoff} E.~S.,  {Bartlett} J.~G.,   {Melin} J.-B.,  2015a, \mn@doi
  [\mnras] {10.1093/mnras/stv605}, \href
  {https://ui.adsabs.harvard.edu/abs/2015MNRAS.450..592R} {450, 592}

\bibitem[\protect\citeauthoryear{{Rozo}, {Rykoff}, {Becker}, {Reddick}  \&
  {Wechsler}}{{Rozo} et~al.}{2015b}]{redmapper4}
{Rozo} E.,  {Rykoff} E.~S.,  {Becker} M.,  {Reddick} R.~M.,   {Wechsler} R.~H.,
   2015b, \mn@doi [\mnras] {10.1093/mnras/stv1560}, \href
  {https://ui.adsabs.harvard.edu/abs/2015MNRAS.453...38R} {453, 38}

\bibitem[\protect\citeauthoryear{{Rykoff} et~al.,}{{Rykoff}
  et~al.}{2014}]{redmapper1}
{Rykoff} E.~S.,  et~al., 2014, \mn@doi [\apj] {10.1088/0004-637X/785/2/104},
  \href {https://ui.adsabs.harvard.edu/abs/2014ApJ...785..104R} {785, 104}

\bibitem[\protect\citeauthoryear{{Schellenberger} \&
  {Reiprich}}{{Schellenberger} \& {Reiprich}}{2017}]{schellenberger2017}
{Schellenberger} G.,  {Reiprich} T.~H.,  2017, \mn@doi [\mnras]
  {10.1093/mnras/stx1022}, \href
  {https://ui.adsabs.harvard.edu/abs/2017MNRAS.469.3738S} {469, 3738}

\bibitem[\protect\citeauthoryear{{Schrabback} et~al.,}{{Schrabback}
  et~al.}{2018}]{Schrabback2018}
{Schrabback} T.,  et~al., 2018, \mn@doi [\mnras] {10.1093/mnras/stx2666}, \href
  {https://ui.adsabs.harvard.edu/abs/2018MNRAS.474.2635S} {474, 2635}

\bibitem[\protect\citeauthoryear{{Simet}, {McClintock}, {Mandelbaum}, {Rozo},
  {Rykoff}, {Sheldon}  \& {Wechsler}}{{Simet} et~al.}{2017}]{Simet2017}
{Simet} M.,  {McClintock} T.,  {Mandelbaum} R.,  {Rozo} E.,  {Rykoff} E.,
  {Sheldon} E.,   {Wechsler} R.~H.,  2017, \mn@doi [\mnras]
  {10.1093/mnras/stw3250}, \href
  {https://ui.adsabs.harvard.edu/abs/2017MNRAS.466.3103S} {466, 3103}

\bibitem[\protect\citeauthoryear{{Sohn}, {Geller}, {Rines}, {Hwang}, {Utsumi}
  \& {Diaferio}}{{Sohn} et~al.}{2018}]{Sohn1712.00872}
{Sohn} J.,  {Geller} M.~J.,  {Rines} K.~J.,  {Hwang} H.~S.,  {Utsumi} Y.,
  {Diaferio} A.,  2018, \mn@doi [\apj] {10.3847/1538-4357/aab20b}, \href
  {https://ui.adsabs.harvard.edu/abs/2018ApJ...856..172S} {856, 172}

\bibitem[\protect\citeauthoryear{{Sohn}, {Geller}, {Hwang}, {Fabricant},
  {Moran}  \& {Utsumi}}{{Sohn} et~al.}{2021}]{Sohn2010.05817}
{Sohn} J.,  {Geller} M.~J.,  {Hwang} H.~S.,  {Fabricant} D.~G.,  {Moran} S.~M.,
    {Utsumi} Y.,  2021, \mn@doi [\apj] {10.3847/1538-4357/abd9be}, \href
  {https://ui.adsabs.harvard.edu/abs/2021ApJ...909..129S} {909, 129}

\bibitem[\protect\citeauthoryear{{Sunayama} et~al.,}{{Sunayama}
  et~al.}{2020}]{Sunayama2020}
{Sunayama} T.,  et~al., 2020, \mn@doi [\mnras] {10.1093/mnras/staa1646}, \href
  {https://ui.adsabs.harvard.edu/abs/2020MNRAS.496.4468S} {496, 4468}

\bibitem[\protect\citeauthoryear{{To} et~al.,}{{To} et~al.}{2021a}]{to2020}
{To} C.,  et~al., 2021a, \mn@doi [\prl] {10.1103/PhysRevLett.126.141301}, \href
  {https://ui.adsabs.harvard.edu/abs/2021PhRvL.126n1301T} {126, 141301}

\bibitem[\protect\citeauthoryear{{To} et~al.,}{{To} et~al.}{2021b}]{to2020a}
{To} C.-H.,  et~al., 2021b, \mn@doi [\mnras] {10.1093/mnras/stab239}, \href
  {https://ui.adsabs.harvard.edu/abs/2021MNRAS.502.4093T} {502, 4093}

\bibitem[\protect\citeauthoryear{{Tomooka}, {Rozo}, {Wagoner}, {Aung}, {Nagai}
  \& {Safonova}}{{Tomooka} et~al.}{2020}]{Tomooka2020}
{Tomooka} P.,  {Rozo} E.,  {Wagoner} E.~L.,  {Aung} H.,  {Nagai} D.,
  {Safonova} S.,  2020, arXiv e-prints, \href
  {https://ui.adsabs.harvard.edu/abs/2020arXiv200311555T} {p. arXiv:2003.11555}

\bibitem[\protect\citeauthoryear{Vikhlinin et~al.,}{Vikhlinin
  et~al.}{2009}]{Vikhlinin_2009}
Vikhlinin A.,  et~al., 2009, \mn@doi [The Astrophysical Journal]
  {10.1088/0004-637x/692/2/1060}, 692, 1060

\bibitem[\protect\citeauthoryear{{Wagoner}, {Rozo}, {Aung}  \&
  {Nagai}}{{Wagoner} et~al.}{2021}]{Wagoner2020}
{Wagoner} E.~L.,  {Rozo} E.,  {Aung} H.,   {Nagai} D.,  2021, \mn@doi [\mnras]
  {10.1093/mnras/stab1012}, \href
  {https://ui.adsabs.harvard.edu/abs/2021MNRAS.504.1619W} {504, 1619}

\bibitem[\protect\citeauthoryear{{Wu}, {Rozo}  \& {Wechsler}}{{Wu}
  et~al.}{2010}]{wu2010}
{Wu} H.-Y.,  {Rozo} E.,   {Wechsler} R.~H.,  2010, \mn@doi [\apj]
  {10.1088/0004-637X/713/2/1207}, \href
  {https://ui.adsabs.harvard.edu/abs/2010ApJ...713.1207W} {713, 1207}

\bibitem[\protect\citeauthoryear{{Yanny} et~al.,}{{Yanny}
  et~al.}{2009}]{segue1}
{Yanny} B.,  et~al., 2009, \mn@doi [\aj] {10.1088/0004-6256/137/5/4377}, \href
  {https://ui.adsabs.harvard.edu/abs/2009AJ....137.4377Y} {137, 4377}

\bibitem[\protect\citeauthoryear{{von der Linden} et~al.,}{{von der Linden}
  et~al.}{2014}]{WTGI}
{von der Linden} A.,  et~al., 2014, \mn@doi [\mnras] {10.1093/mnras/stt1945},
  \href {https://ui.adsabs.harvard.edu/abs/2014MNRAS.439....2V} {439, 2}

\makeatother
\end{thebibliography}
\immediate\write16{No file \jobname.bbl.} 



\appendix
\section{Validation of fiducial double-Gaussian model}
\label{app:validation}
\begin{figure}
  \centering
  \includegraphics[scale=0.40]{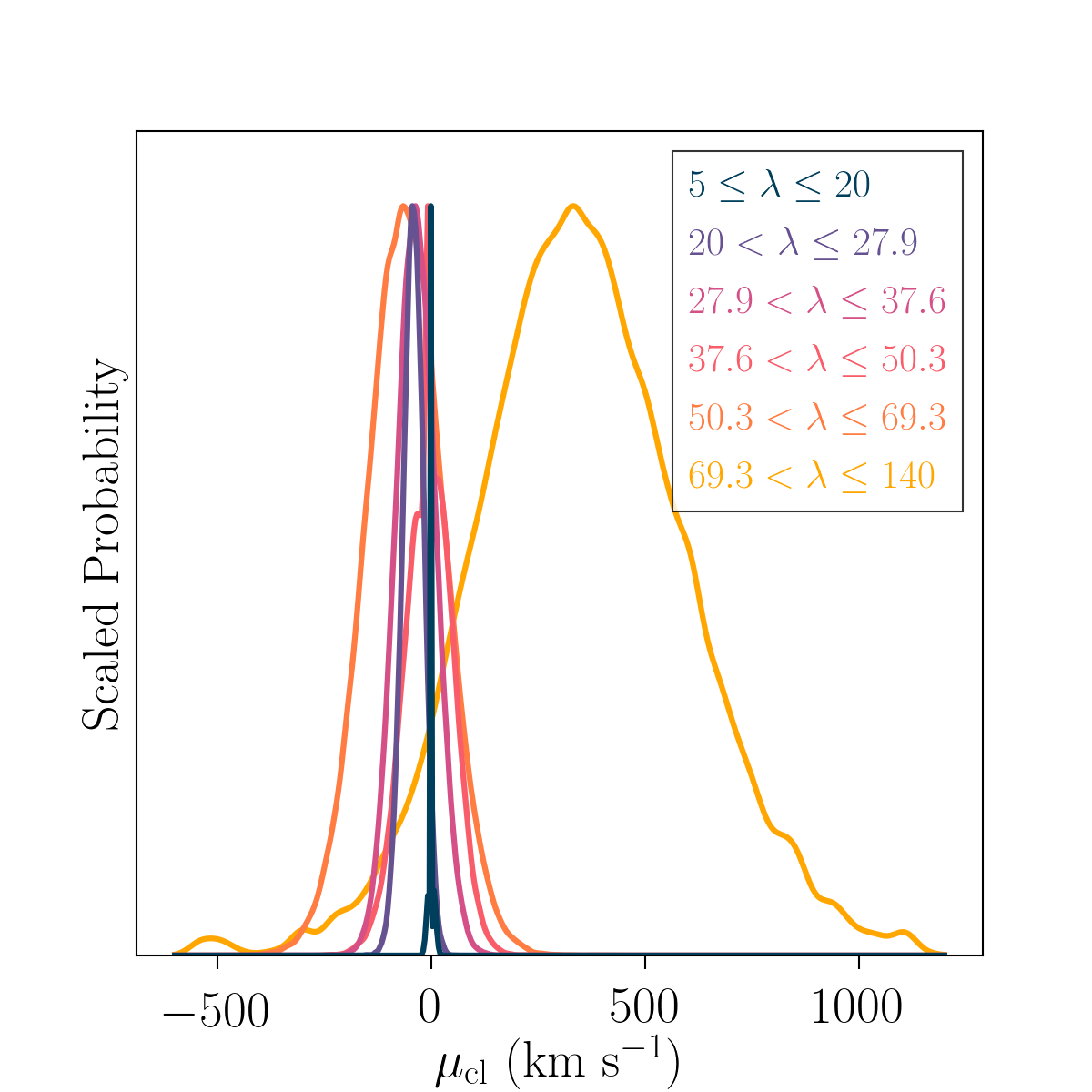}
  \caption{
     Posterior probability distributions describing the $\mu_{\text{cl}}$ values in the six richness bins when included as independent free parameters in the fits. In all cases the mean redshift offsets are consistent with zero. The results validate our assumption of $\mu_{\text{cl}}=0$ in the fiducial model fit. Here uncertainties are determined by bootstrap resampling the cluster sample.
  }
  \label{fig:sdss_cl_mu_vs_richness}
\end{figure}

Our fiducial double-Gaussian model assumes that the virialized cluster velocity component has a mean redshift offset of zero from the BCGs, i.e. $\mu_{\text{cl}}=0$. Fig. \ref{fig:sdss_cl_mu_vs_richness} shows the posterior probability distributions for the $\mu_{\text{cl}}$ values for the six richness bins when included as independent free parameters in the fits. In all cases the mean redshift offsets are consistent with zero.  

\begin{figure}
   \centering
   \includegraphics[width=\linewidth]{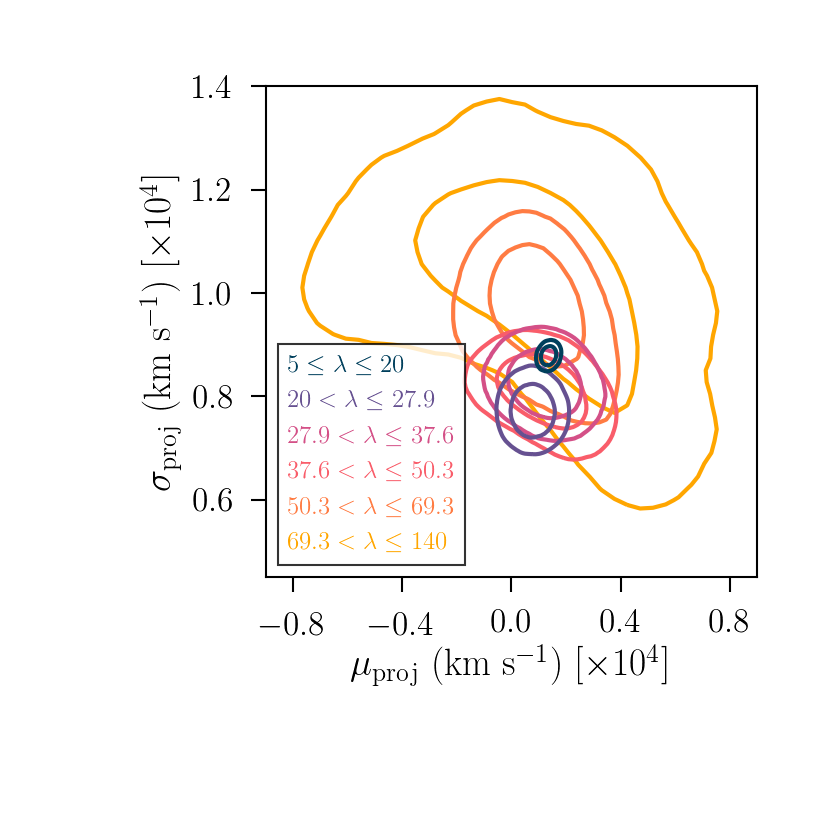}
   \caption{Joint posterior probability distributions for $\mu_{\text{proj}}$ and $\sigma_{\text{proj}}$, describing the mean velocity and velocity width of the projected components, when included as independent free parameters for each richness bin. Uncertainties are determined by bootstrap resampling the cluster population. The consistency of the results across all richness bins justifies the assumption of a common velocity width and zero velocity offset for the projected component in the fiducial analysis.}\label{fig:sdss_bkg_params_vs_richness}
\end{figure}

Our fiducial model additionally assumes that the velocity widths of the projected components, $\sigma_{\text{proj}}$ are consistent with a common value. Fig.  \ref{fig:sdss_bkg_params_vs_richness} shows the joint constraints on $\mu_{\text{proj}}$ and $\sigma_{\text{proj}}$ for the six richness bins, when included as free parameters in the fits. The results again validate the fiducial model assumptions. 

\begin{figure}
  \centering
  \includegraphics[scale=0.40]{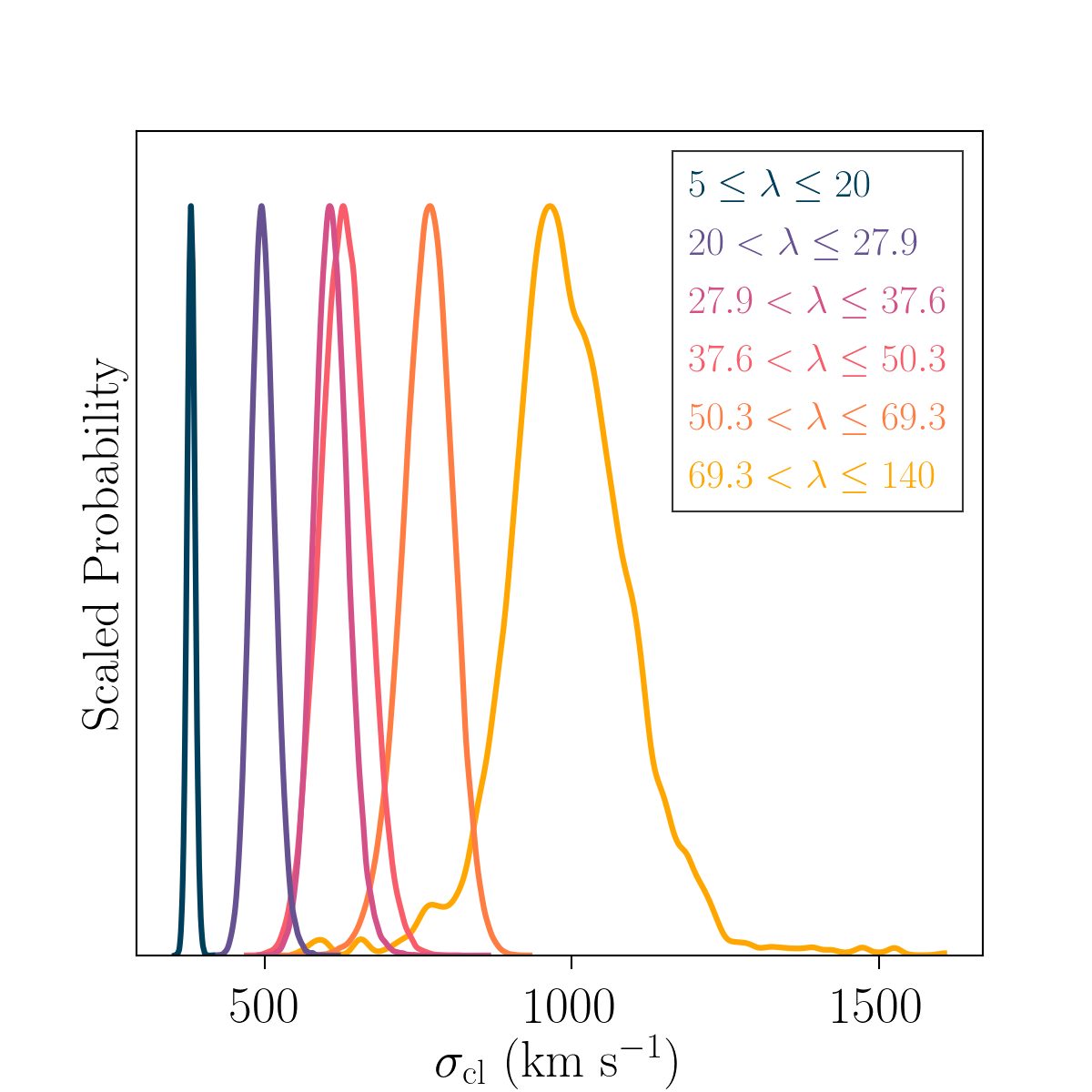}
  \caption{
     Posterior probability distributions describing the $\sigma_{\text{cl}}$ values in the six richness bins, included as independent free parameters in the fits. Here uncertainties are determined by bootstrap resampling the cluster population.
  }
  \label{fig:sdss_cl_sig_vs_richness}
\end{figure}

As a final qualitative validation, Fig. \ref{fig:sdss_cl_sig_vs_richness} shows the posterior distributions for the widths of the observed cluster velocity components, $\sigma_{{\rm cl,j}}$, in the six richness bins. As expected, the velocity width increases smoothly as a function of $\lambdaobs$.

Our final best fitting model parameters are shown in Table \ref{tab:parameters}.

\begin{table*}
\begin{center}
\begin{tabular}[h]{l c c c c c c } 

Richness & 5 -- 20 & 20 -- 27.9 &  27.9 -- 37.6 & 37.6 -- 50.3 & 50.3 -- 69.3 & 69.3 -- 140\\ 
\hline
\hline
$\sigma_\mathrm{cl}$ (km s$^{-1}$) & $379\pm6$ &$503\pm20$& $614\pm30$ & $634\pm40$& $770\pm40$& $1060\pm139$\\

$f_\mathrm{proj}$ & $0.392 \pm 0.007$ & $0.265\pm0.021$& $0.243\pm0.025$& $0.251\pm0.045$& $0.137\pm0.024$& $0.083\pm0.025$\\

$b_{\lambda}$ (data) & $0.157\pm 0.006$ & 
$0.126\pm0.020$&$0.127\pm0.024$& $0.160\pm0.042$& $0.078\pm0.022$&$0.033\pm0.030$\\

$b_{\lambda}$ (\textsc{Buzzard}) & $0.080\pm0.010$& $0.095\pm0.015$&$0.070\pm0.020$& $0.082\pm0.031$& $0.070\pm0.027$&$0.059\pm0.036$\\

$b_{\lambda}$ (Costanzi et al.) &$0.1204\pm0.0003$ & $0.1239\pm0.0009$ &$0.1085\pm0.0012$ &$0.0952\pm0.0016$ &$0.0853\pm0.0020$ &$0.0824\pm0.0024$\\

\rule{0pt}{4ex} 
\rule{0pt}{4ex} 

\end{tabular}\caption{Best fitting parameters for the projection effects models and richness bias described in this work. The cluster component mean $\mu_{\mathrm{cl}}$ is fixed to zero for all bins in the fiducial fit and the projection component parameters are fit jointly to all bins and found to be $\sigma_\mathrm{proj} = 8689 \pm 1074$ km s$^{-1}$ and $\mu_\mathrm{proj} = 1299 \pm 161$ km s$^{-1}$. }\label{tab:parameters}
\end{center}
\end{table*}

\section{Completeness of spectroscopic coverage}
\label{app:completeness}
\begin{figure}
  \centering
  \includegraphics[scale=0.30]{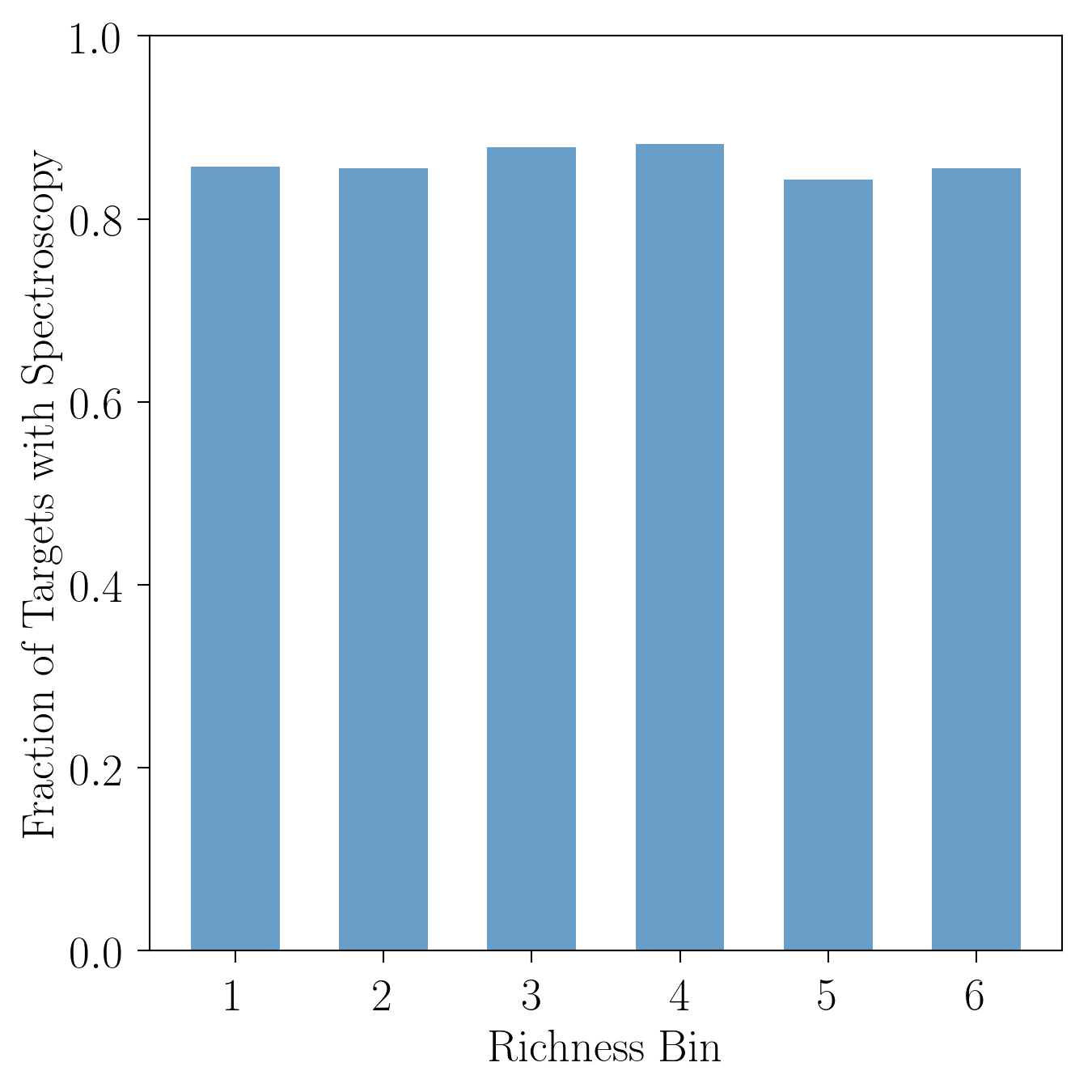}
  \caption{
     Spectroscopic completeness of the selected cluster member sample from the archival SDSS data used in this study. As shown here, the spectroscopic coverage of the selected cluster member sample is uniform across richness bins.
  }
  \label{fig:sdss_data_completeness}
\end{figure}

Chief among the data requirements for this study is the existence of spectroscopic coverage of cluster members samples that is representative of the cluster member population. In other words, we require samples of cluster member galaxies with spectroscopic redshift measurements whose distribution in any observed parameter is unbiased relative to the cluster member population as a whole. The primary selection function we apply to achieve this with the archival data used in this study is the fiducial luminosity cut $L\geq0.55\lstar$. This cut yields [37, 37, 38, 37, 39, and 39] percent of the members of the selected clusters in the six cluster richness bins, respectively. After this cut, the completeness of the spectroscopic coverage is nearly uniform ranging from approximately 84 to 88 percent, as shown in Figure \ref{fig:sdss_data_completeness}.

\section*{Affiliations}
$^{1}$ Department of Physics, Stanford University, 382 Via Pueblo Mall, Stanford, CA 94305, USA\\
$^{2}$ SLAC National Accelerator Laboratory, Menlo Park, CA 94025, USA\\
$^{3}$ Kavli Institute for Particle Astrophysics \& Cosmology, P. O. Box 2450, Stanford University, Stanford, CA 94305, USA\\
$^{4}$INAF-Osservatorio Astronomico di Trieste, via G. B. Tiepolo 11, I-34143 Trieste, Italy\\
$^{5}$Astronomy Unit, Department of Physics, University of Trieste, via Tiepolo 11, I-34131 Trieste, Italy\\
$^{6}$Institute for Fundamental Physics of the Universe, Via Beirut 2, 34014 Trieste, Italy\\
$^{7}$Berkeley Center for Cosmological Physics, University of California, Berkeley, CA 94720, USA\\
$^{8}$Lawrence Berkeley National Laboratory ,1 Cyclotron Road, Berkeley, CA 94720, US\\
$^{9}$Fermi National Accelerator Laboratory, P. O. Box 500, Batavia, IL 60510, USA\\
$^{10}$Department of Physics, University of California, 1156 High St, Santa Cruz, CA 95064, USA\\
$^{11}$Santa Cruz Institute for Particle Physics, University of California, 1156 High St, Santa Cruz, CA 95064, USA\\
$^{12}$Department of Physics, University of Arizona, 1118 E. Fourth Street, Tucson, AZ, 85721, USA\\
$^{13}$Gemini Observatory/NSF's NOIRLab, Casilla 603, La Serena, Chile\\


\bsp	
\label{lastpage}
\end{document}